\documentclass[aps,prb,twocolumn,superscriptaddress,floatfix,showkeys]{revtex4-2}

\usepackage[english]{babel}
\usepackage[dvips]{graphicx}
\usepackage{graphics}
\usepackage{amsfonts}
\usepackage{amsmath}
\usepackage{amssymb}
\usepackage{exscale}
\usepackage{eufrak}
\usepackage{afterpage}
\usepackage{upgreek}
\usepackage{color}	
\usepackage{braket}	
\usepackage{multirow}	
\usepackage{enumitem}
\usepackage{balance}

\usepackage{epstopdf}
\usepackage{miller}	
\usepackage{units}
\usepackage{stmaryrd} 
\definecolor{OliveGreen}{rgb}{0,0.6,0}
\usepackage{hyperref}

\usepackage[normalem]{ulem}    

\newcommand{\me}{m_\mathrm{e}}
\newcommand{\mheb}{m_\mathrm{HEB}}
\newcommand{\mleb}{m_\mathrm{LEB}}
\newcommand{\mhhb}{m_\mathrm{HHB}}
\newcommand{\mlhb}{m_\mathrm{LHB}}
\newcommand{\Vhyb}{V_\mathrm{hyb}}
\newcommand{\Vfold}{V_\mathrm{fold}}
\newcommand{\Vheavy}{V_\mathrm{heavy}}
\newcommand{\Vlight}{V_\mathrm{light}}

\begin{document}
\title{From hidden-order to antiferromagnetism:\\
	   electronic structure changes in Fe-doped URu$_{2}$Si$_{2}$}

\author{Emmanouil Frantzeskakis}
\thanks{emmanouil.frantzeskakis@universite-paris-saclay.fr}
\affiliation{Universit\'e Paris-Saclay, CNRS,  Institut des Sciences Mol\'eculaires d'Orsay, 
			91405, Orsay, France}
						
\author{Ji Dai}
\affiliation{Universit\'e Paris-Saclay, CNRS,  Institut des Sciences Mol\'eculaires d'Orsay, 
			91405, Orsay, France}
\affiliation{Institute of Physics and Lausanne Centre for Ultrafast Science (LACUS), 
			\'Ecole Polytechnique F\'ed\'erale de Lausanne, CH-1015 Lausanne, Switzerland}

\author{C\'edric Bareille}
\affiliation{Institute for Solid State Physics, University of Tokyo, Chiba 277-8581, Japan}

\author{Tobias C. R\"odel}
\affiliation{Universit\'e Paris-Saclay, CNRS,  Institut des Sciences Mol\'eculaires d'Orsay, 
			91405, Orsay, France}
\affiliation{Synchrotron SOLEIL, L'Orme des Merisiers, Saint-Aubin-BP48, 91192 Gif-sur-Yvette, France}

\author{Monika G\"uttler}
\affiliation{Universit\'e Paris-Saclay, CNRS,  Institut des Sciences Mol\'eculaires d'Orsay, 
			91405, Orsay, France}
\affiliation{Institut f\"ur Festk\"orperphysik, Technische Universit\"at Dresden, 
			D-01062 Dresden, Germany}

\author{Sheng~Ran}
\affiliation{Department of Physics, University of California-San Diego, 9500 Gilman Drive, 
		 La Jolla, California 92093, USA}
\affiliation{Department of Physics, University of Maryland, College Park, Maryland 20742, USA}

\author{Noravee~Kanchanavatee}
\affiliation{Department of Physics, University of California-San Diego, 9500 Gilman Drive, 
		 La Jolla, California 92093, USA}
\affiliation{Department of Physics, Chulalongkorn University, Pathumwan 10330, Thailand}

\author{Kevin~Huang}
\affiliation{Department of Physics, University of California-San Diego, 9500 Gilman Drive, 
		 La Jolla, California 92093, USA}
\affiliation{Materials Science Division, Lawrence Livermore National Laboratory, 
		 Livermore, CA, 94550, USA}

\author{Naveen~Pouse}
\affiliation{Department of Physics, University of California-San Diego, 9500 Gilman Drive, 
		 La Jolla, California 92093, USA}

\author{Christian~T.~Wolowiec}
\affiliation{Department of Physics, University of California-San Diego, 9500 Gilman Drive, 
		 La Jolla, California 92093, USA}

\author{Emile~D.~L.~Rienks}
\affiliation{Helmholtz-Zentrum Berlin f\"ur Materialien und Energie, 
		 Elektronenspeicherring BESSY II, Albert-Einstein-Strasse 15, 12489 Berlin, Germany}

\author{Pascal~Lejay}
\affiliation{Institut N\'eel, CNRS and Universit\'e Grenoble Alpes, F-38042 Grenoble, France}

\author{Franck~Fortuna}
\affiliation{Universit\'e Paris-Saclay, CNRS,  Institut des Sciences Mol\'eculaires d'Orsay, 
			91405, Orsay, France}

\author{M.~Brian~Maple}
\thanks{mbmaple@ucsd.edu}
\affiliation{Department of Physics, University of California-San Diego, 9500 Gilman Drive, 
		 La Jolla, California 92093, USA}

\author{Andr\'esF.~Santander-Syro}
\thanks{andres.santander-syro@universite-paris-saclay.fr}
\affiliation{Universit\'e Paris-Saclay, CNRS,  Institut des Sciences Mol\'eculaires d'Orsay, 
			91405, Orsay, France}

\begin{abstract}
	In matter, any spontaneous symmetry breaking induces a phase transition 
	characterized by an order parameter, such as the magnetization vector in ferromagnets, 
	or a macroscopic many-electron wave-function in superconductors. 
	Phase transitions with unknown order parameter are rare but extremely appealing, 
	as they may lead to novel physics. 
	An emblematic, and still unsolved, example is the transition of the heavy fermion compound 
	URu$_2$Si$_2$ (URS) into the so-called hidden-order (HO) phase 
	when the temperature drops below $T_0 = 17.5$K. 
	Here we show that the interaction between the heavy fermion and the conduction band states 
	near the Fermi level has a key role in the emergence of the HO phase. 
	Using angle resolved photoemission spectroscopy, we find that while the Fermi surfaces 
	of the HO and of a neighboring antiferromagnetic (AFM) phase 
	of well-defined order parameter have the same topography; 
	they differ in the size of some, but not all, of their electron pockets. 
	Such a non-rigid change of the electronic structure indicates 
	that a change in the interaction strength 
	between states near the Fermi level is a crucial ingredient for
	the HO-to-AFM phase transition.
\end{abstract}

\url{www.pnas.org/cgi/doi/10.1073/pnas.2020750118}


\maketitle

The transition of URu$_{2}$Si$_{2}$ from a high-temperature paramagnetic (PM) phase 
to the HO phase below $T_{0}$ is accompanied 
by anomalies in specific heat~\cite{Maple1986, Palstra1985, Schlabitz1986}, 
electrical resistivity~\cite{Maple1986, Schlabitz1986}, thermal expansion~\cite{deVisser1986}
and magnetic susceptibility~\cite{Palstra1985, Schlabitz1986} 
that are all typical of magnetic ordering. 
However, the small associated antiferromagnetic (AFM) moment~\cite{Broholm1987} 
is insufficient to explain the large entropy loss,
and was shown to be of extrinsic origin~\cite{Matsuda2001}.
Inelastic neutron scattering (INS) experiments revealed gapped magnetic excitations 
below $T_{0}$ at commensurate and incommensurate 
wave-vectors~\cite{Wiebe2007, Villaume2008, Butch2015},
while an instability and partial gapping of the Fermi surface was observed 
by angle-resolved photoemission spectroscopy (ARPES)~\cite{Denlinger2001, Santander2009, 
Yoshida2012, Yoshida2013, Meng2013, Boariu2013, Chatterjee2013} 
and scanning tunneling microscopy/spectroscopy (STM/STS)~\cite{Schmidt2010, Aynajian2010}. 
More recently, high-resolution low-temperature ARPES experiments imaged
the Fermi surface (FS) reconstruction across the hidden-order transition,
unveiling the nesting vectors between Fermi sheets associated
with the gapped magnetic excitations seen in INS experiments~\cite{Meng2013, Bareille2014},
and quantitatively explaining, from the changes in Fermi surface size and quasiparticle mass,
the large entropy loss in the hidden-order phase~\cite{Bareille2014}.
Nonetheless, the nature of the HO parameter is still 
hotly debated~\cite{Haule2009,Elgazzar2009,Mydosh2011,Mydosh2014}.

The HO phase is furthermore unstable above a temperature-dependent 
critical pressure, of about $0.7$~GPa at $T=0$, 
at which it undergoes a first-order transition into a large moment AFM phase
where the value of the magnetic moment per U atom exhibits a sharp increase, 
by a factor of 10-50~\cite{McElfresh1987,Amitsuka1999,Matsuda2001,Motoyama2003,Jeffries2007,
Butch2010,Bourdarot2011,Knafo2020}.
When the system crosses the HO$\rightarrow$AFM phase boundary, 
the characteristic magnetic excitations of the HO phase are either suppressed  
or modified ~\cite{Villaume2008, Williams2017}, while resistivity and specific heat measurements 
suggest that the partial gapping of the Fermi surface is enhanced~\cite{McElfresh1987, Jeffries2007}.

As the AFM phase has a well defined order parameter, studying the evolution of the 
electronic structure across the HO/AFM transition would help 
develop an understanding of the HO state. 
So far the experimental determination of the Fermi surface 
by means of Shubnikov de Haas (SdH) oscillations only showed minor changes 
across the HO$\rightarrow$AFM phase boundary~\cite{Hassinger2010}.
Here, we take advantage of the HO/AFM transition induced by chemical pressure 
in URu$_{2}$Si$_{2}$, through the partial substitution of Ru 
with Fe~\cite{Kanchanavatee2011, Das2015, Butch2016, Kung2016, Ran2016},
to directly probe its electronic structure in the AFM phase using ARPES. 
As we shall see, our results reveal that changes in the Ru~$4d$~-U~$5f$ hybridisation 
across the HO-AFM phase boundary seem essential for a better understanding of the HO state.

ARPES measurements were performed at the UE112-PGM-2b-1$^{3}$ endstation of BESSY~II 
using a hemispherical electron analyser. 
The instrumental resolution varied from 3 to 7~meV according to experimental conditions.
Samples were cleaved in situ at a temperature below 20K,
and measured at temperatures between 1~K and 25~K.
The pressure was at all times lower than $1.0 \times 10^{-10}$~mbar. 
Single crystals of (Fe-doped) URu$_{2}$Si$_{2}$ were grown in a tetra-arc furnace 
using the Czochralski method in an argon atmosphere. 
The quality of the synthesised crystals was confirmed by X-ray diffraction measurements. 
Sample pieces were oriented by means of Laue diffraction.
The Supplementary Information (SI.\ref{SI:URS-Xtals}, SI.\ref{SI:ARPES-details} 
and SI.\ref{SI:3D-mapping}) 
provides complete technical details
about the crystal growth, characterization, and ARPES measurements.

\section*{Results \& Discussion}
\begin{figure}[!t]
\centering
	\includegraphics[clip, width=\linewidth]{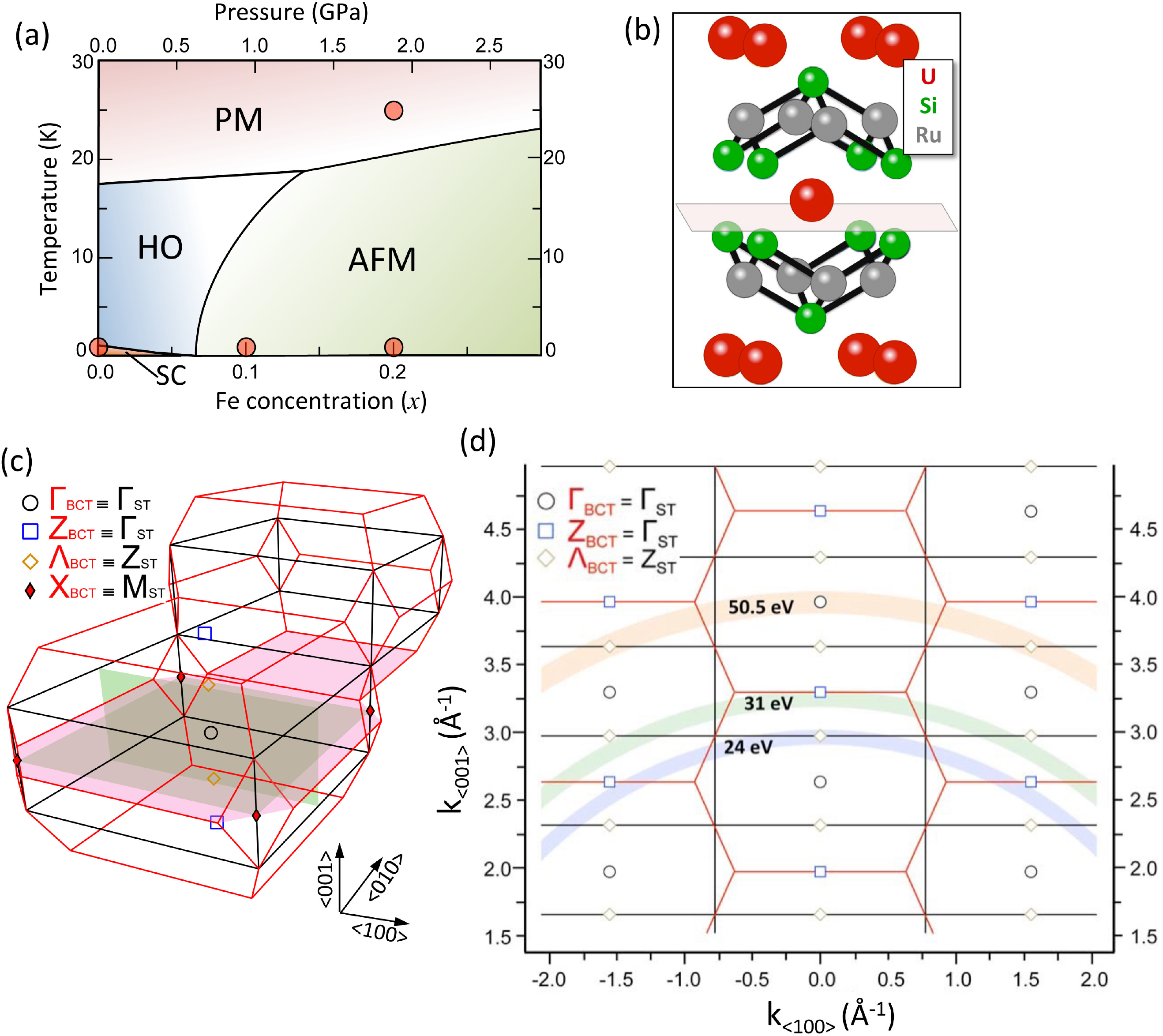}
    \caption{
        	 (a)~Temperature phase diagram of URu$_{2-x}$Fe$_{x}$Si$_{2}$ 
        	 as a function of Fe concentration and equivalent chemical pressure 
        	 as determined by means of electrical resistivity, specific heat, magnetization, 
        	 thermal expansion and neutron diffraction 
        	 measurements~\cite{Ran2016, Das2015, Kung2016, Kanchanavatee2011}. 
        	 The markers denote the different compositions/temperatures 
        	 studied in the present work.
        	 Henceforth, blue and green hues will be used to represent ARPES data
        	 in the HO and AFM states, respectively.
        	 (b)~Schematic unit cell, showing a natural Si-terminated cleavage plane.
        	 (c)~Bulk body-centered tetragonal (BCT, red lines, PM phase) and simple tetragonal 
        	 (ST, black lines, HO and AFM phases) Brillouin zones. 
        	 The typical in-plane (red) and out-of-plane (green) measurement planes,
        	 together with the high-symmetry 
        	 $\Gamma$, $\textmd{Z}$ and $\Lambda$ points,
        	 are also shown. 
        	 Note that, in the folded ST Brillouin zone,
        	 The $\Gamma$ and $\textmd{Z}$ points of the BCT zone become equivalent, and
        	 $\textmd{Z}_\textmd{ST} = \Lambda_\textmd{BCT}$. 
        	 (d)~Reciprocal $k_{<100>}-k_{<001>}$ plane showing the spherical caps in $k$-space 
        	 probed with 50.5~eV, 31~eV and 24~eV photons for different emission angles 
        	 of the photoelectrons, within the free-electron model of the 
        	 photoemission final states.
        	 Black (red) lines denote the borders of the simple tetragonal 
        	 (body-centered tetragonal) Brillouin zone. 
        	 Damping of the final states results in broadening 
        	 of the surface-perpendicular wave-vector. 
        	 The corresponding uncertainty ($\delta k_{\bot}$),
        	 represented by the widths of the arcs, 
        	 is related to the photoelectron escape depth ($\lambda$) as 
        	 $\delta k_{\bot}$ = $1/\lambda$~\cite{Strocov2003}.
        	 The inner potential has been set to 
        	 13~eV~\cite{Denlinger2001,Santander2009,Boariu2013,Bareille2014}. 
        }
	\label{fig1}
\end{figure}

Fig.~\ref{fig1}(a) shows the phase diagram of URu$_{2-x}$Fe$_{x}$Si$_{2}$ 
as a function of Fe concentration ($x$) and of the associated chemical pressure.
The quantum-critical transition at $T = 0$ from the HO to the AFM phase occurs at 
$x_c \approx 0.07$. 
Our samples (orange markers) are characteristic of the HO, AFM and PM phases. 
This study will focus on changes across the HO/AFM phase boundary. 
As schematized in Fig.~\ref{fig1}(b), henceforth 
we will characterize the electronic structure of Fe-doped URu$_2$Si$_2$(001) surfaces 
with a Si termination layer, which corresponds to a buried bulk-like U layer, 
as the bulk-derived heavy bands of U~$5f$ origin are a key feature of the low energy
electronic structure of URu$_2$Si$_2$~\cite{Santander2009, Yoshida2012, Yoshida2013, 
Boariu2013, Chatterjee2013, Bareille2014}.
A detailed comparison with the alternative U-terminated surfaces,
and the stability of cleaved surfaces in UHV, 
is presented in the Supplementary Information 
(SI.\ref{SI:ARPES-SurfaceTerm} and SI.\ref{SI:Robustness-cleave-surface}).

Fig.~\ref{fig1}(c) presents the 3D body-centered tetragonal (BCT, PM phase, red lines)
and simple-tetragonal (ST, HO and AFM phases, black lines) Brillouin zone (BZ) of URS, 
and highlights the typically measured planes. 
Fig.~\ref{fig1}(d) illustrates how reciprocal space is probed
using photon energies of $50.5$~eV, $31$~eV and $24$~eV. 
At normal emission ($k_{<100>}=0$) these energies correspond to bulk
$\Gamma$, $\textmd{Z}$ and $\Lambda$ high-symmetry points of the BCT Brillouin zone. 
Off-normal emission, in the neighboring Brillouin zones,
the same photon energies probe the states near the $\Lambda$ ($50.5$~eV and $31$~eV)
and $\textmd{Z}$ ($24$~eV) high-symmetry points. 

\begin{figure*}[!t]
\centering
	\includegraphics[clip, width=\textwidth]{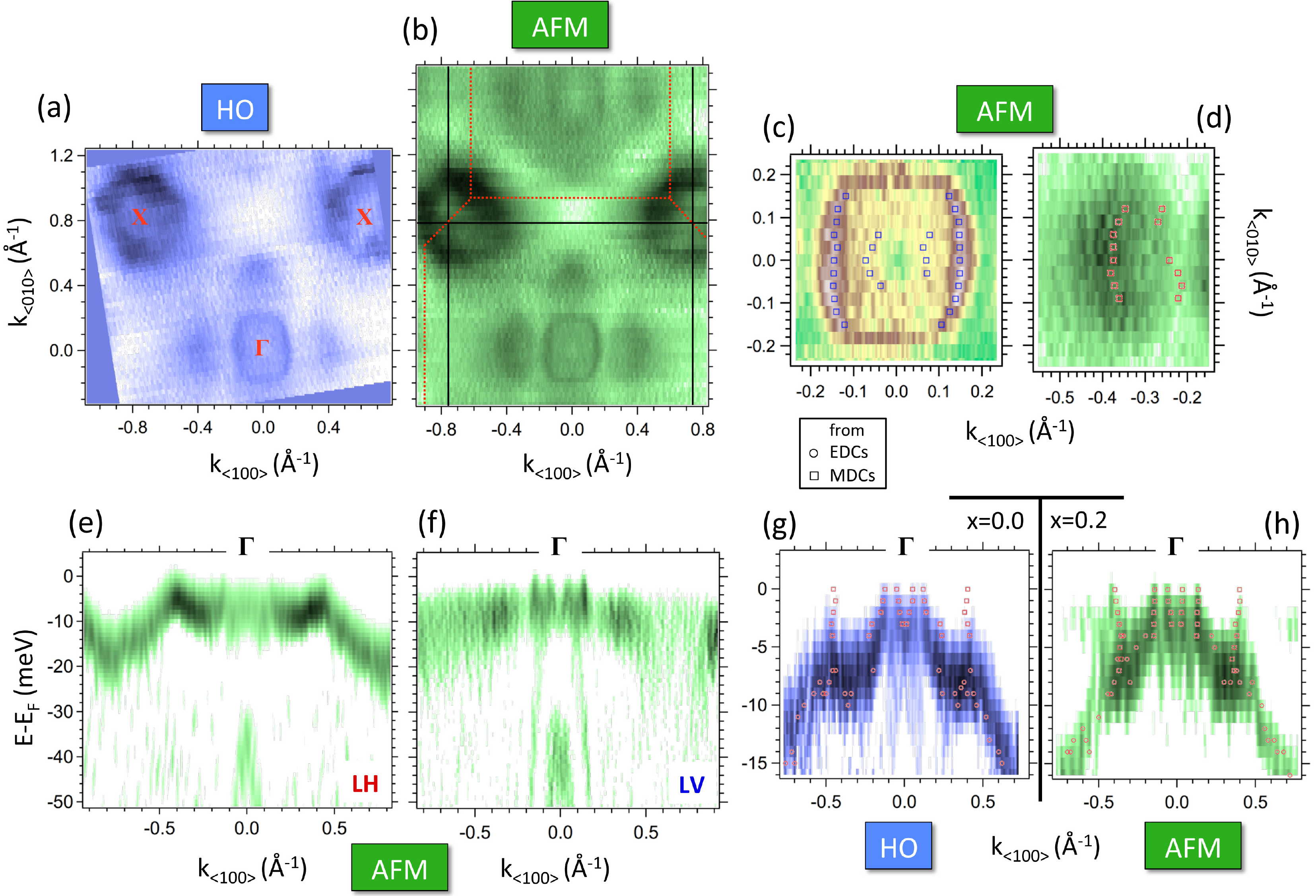}
    \caption{
        	 (a,~b) In-plane Fermi surface maps of pure (blue hues) and Fe-doped (green hues)
        	 URu$_2$Si$_2$ in the HO and AFM phases, respectively. 
        	 Red and black lines denote the BCT and ST Brillouin zones.
        	 (c,~d) Zoom on two characteristic common features of the HO and AFM states,
        	 here measured on a URu$_{1.9}$Fe$_{0.1}$Si$_2$ sample in the AFM state: 
        	 the pockets around $\Gamma$, and the off-centered Fermi-petals, respectively. 
        	 (e,~f) Near-$E_{\textmd{F}}$ energy dispersion of Fe-doped URu$_{2}$Si$_{2}$ 
        	 using photons with LH or LV polarisation, 
        	 as to enhance $f$- and $d$-derived features, respectively. 
        	 (g,~h) High-resolution ARPES energy-momentum dispersions 
        	 around $\Gamma$ along $k_{<100>}$ in, respectively, 
        	 the HO phase of URu$_{2}$Si$_{2}$ and the AFM phase
        	 of URu$_{1.8}$Fe$_{0.2}$Si$_{2}$. 
        	 Photons with LH polarisation were used.
        	 All data in this figure were measured at 1K with 50.5~eV photons. 
        	 In order to enhance the experimental features, 
        	 the 2D curvature of the ARPES intensity is presented~\cite{Zhang2011}.
        	 The Fermi-surface maps are a superposition of identical measurements 
        	 using photons with LH and LV polarisation,
        	 in order to display all the features with different orbital character.
        	 Markers in panels (c,~d,~g,~h) show the local maxima of the spectral function
        	 extracted from either energy distribution curves (EDCs, circles) or 
        	 momentum distribution curves (MDCs, squares).
         	 }
	\label{fig2}
\end{figure*}

Figs.~\ref{fig2}(a,~b) show the experimental in-plane Fermi surface of pure 
and Fe-doped URu$_2$Si$_{2}$ (Fe-URS) 
in the low-temperature HO and AFM states, respectively.
Their striking similarity demonstrates that the HO and AFM phases
share common nesting vectors, as inferred 
from INS experiments~\cite{Wiebe2007, Villaume2008, Butch2015}
and theoretical calculations~\cite{Elgazzar2009,Ikeda2012},
and directly confirms conclusions from previous 
measurements of extremal Fermi surface contours by means of SdH oscillations, 
which showed only minor differences between the HO and the AFM phases~\cite{Hassinger2010}. 
Specifically, as seen from Figs.~\ref{fig2}(a,~b), 
there are two Fermi sheets centered around $\Gamma$: 
an intense square-like outer contour with $k_F \approx 0.15$~\AA$^{-1}$ along $k_{<100>}$, 
and a circumscribed weaker circular-like inner contour 
--see Fig.~\ref{fig2}(c). 
These contours correspond, respectively, to hole-like and electron-like pockets. 
The pockets are formed by an M-shaped band resulting from the interaction between 
heavy and light states, as discussed in previous works 
on pure URu$_2$Si$_2$~\cite{Bareille2014}, 
and described by means of a phenomenological toy model 
in the Supplementary Information (SI.\ref{SI:Toy-model}, Fig.~\ref{figS7} and Fig.~\ref{figS8}).
At larger $k_{<100>}$ momenta of the order of $0.3-0.4$~\AA$^{-1}$,
there are four off-centered ``Fermi Petals'' 
symmetrically distributed around $\Gamma$~\cite{Bareille2014},
shown in detail in Fig.~\ref{fig2}(d). 
All these Fermi sheets around $\Gamma$ agree well with previous LSDA calculations 
in the AFM phase of URS~\cite{Elgazzar2009,Oppeneer2010}.
Additionally, large electron pockets centered at the corners 
of the ST Brillouin zone (the $\textmd{X}$ points) 
are observed in both the HO and AFM states --Figs.~\ref{fig2}(a,~b).
These pockets arise from the interaction (hybridization) of heavy bands of U~$5f$ character 
with a dispersing hole-like band~\cite{Bareille2014}. 

Figs.~\ref{fig2}(e,~f) present band dispersions of Fe-URS along $k_{<100>}$ 
using photons with linear horizontal (LH) and linear vertical (LV) polarisation, respectively. 
The dispersion of the heavy bands can be best detected using LH photons, 
while a light hole-like band and the M-shaped feature are best probed with LV photons.
The necessity for light with variable polarisation reflects the different orbital character 
of those states, namely the heavy bands of U~$5f$ origin and the light bands of likely Ru~$4d$ origin. 
The phenomenological model presented in the Supplementary Information (SI.\ref{SI:Toy-model})
includes two pairs of bands with very different effective masses 
to mimic the light and heavy bands.
Finally, the hole-like band with a maximum at $E-E_F \approx -30$~meV 
is common to both polarisations and corresponds to a surface state~\cite{Boariu2010}. 

Figs.~\ref{fig2}(g,~h) show a zoom of the near-$E_F$ electronic dispersion around $\Gamma$,
both for the pure and Fe-doped compounds.
One clearly observes the M-shaped band around $k_{<100>}=0$ 
forming the outer square-like hole pocket and inner circular-like electron pocket.
The shallow Fermi petals form above the ``spikes'' of the M-shaped band
at $k_{<100>} \approx 0.4$~\AA$^{-1}$.
Our experimental resolution, which worsens as photon energy increases, 
does not allow us to draw conclusions about
electronic structure changes around $\Gamma$ 
between the pure (HO) and Fe-doped (AFM) compounds. 
On the other hand, as we shall see next, the electronic structure 
around the $\Lambda$ and $\textmd{Z}$ points, 
which are both probed with smaller photon energies, 
show clear changes between the two phases.

\begin{figure*}[!ht]
 	\centering
    \includegraphics[clip, width=\textwidth]{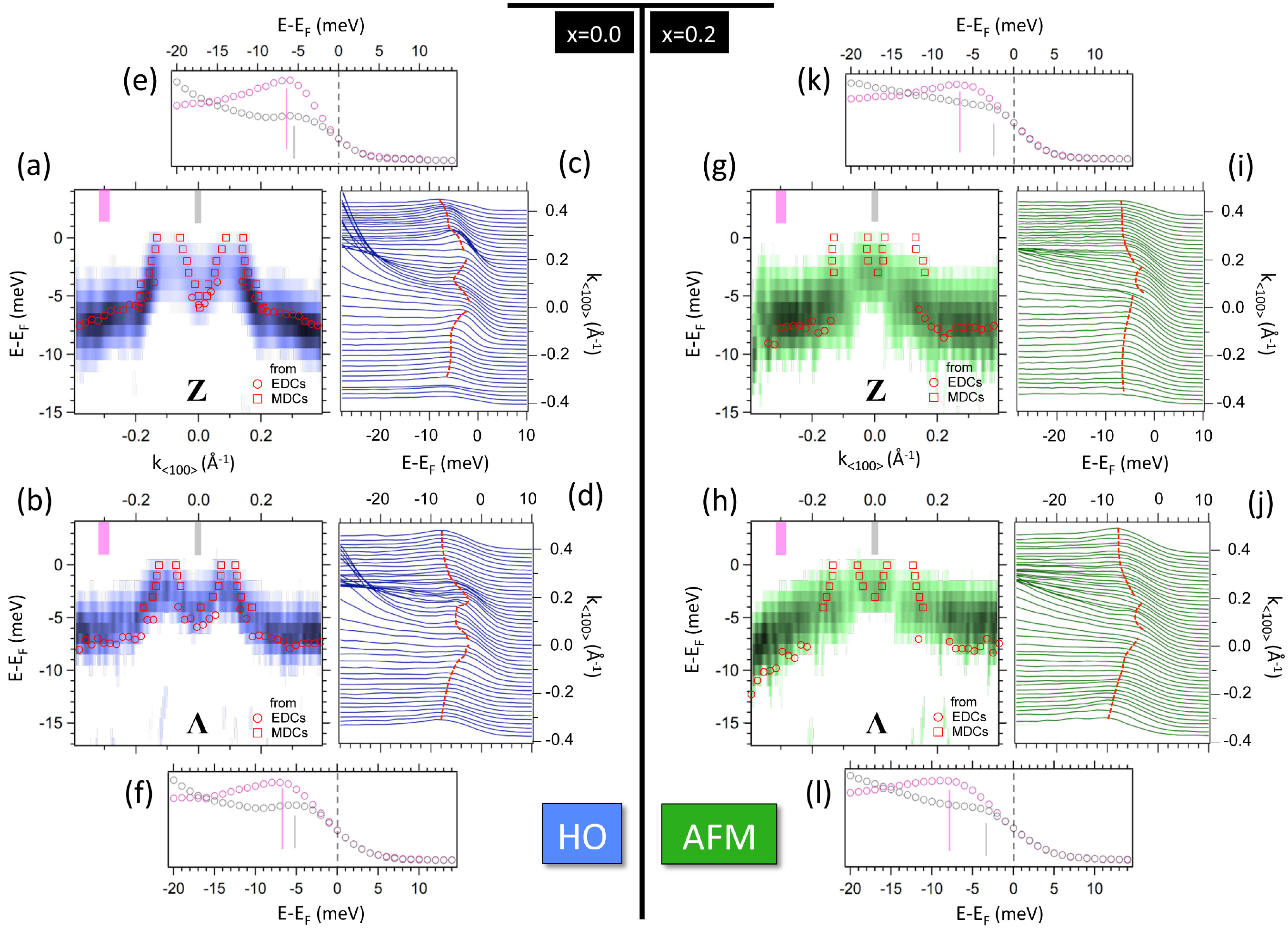}
    \caption{
        	 (a,~b)~High-resolution ARPES energy-momentum maps (2D curvatures~\cite{Zhang2011}) 
        	 in the HO phase (pure URu$_{2}$Si$_{2}$) around the Z and $\Lambda$ points,
        	 respectively. 
        	 Red markers show the local maxima of the spectral function
        	 extracted from either EDCs (circles) or MDCs (squares).
        	 (c~d)~Corresponding EDCs of the raw ARPES data. 
        	 Dashed red lines are guides to the eye.
        	 (e,~f)~EDCs at $k_{<100>}=0$ (gray circles) and 
        	 $k_{<100>}=0.3$\AA$^{-1}$~(magenta circles) around Z and $\Lambda$,
        	 respectively. Thin vertical lines mark the energy of the corresponding EDC peak. 
        	 The EDCs were integrated over the momenta range shown by the thick bars
        	 in panels (a) and (b).
        	 (g-l)~Measurements analogous to (a-f) in the AFM phase 
        	 (URu$_{1.8}$Fe$_{0.2}$Si$_{2}$).
        	 To facilitate comparisons, energy-momentum maps in panels (b) and (h) 
        	 have been interpolated in the energy direction before taking the curvature. 
        	 The original energy step was 2~meV.
        	 Data around Z and $\Lambda$ have been acquired using, respectively, 
        	 31~eV and 24~eV photons with LH polarisation. 
        	 All data were measured at 1~K.
        	}
	\label{fig3}
\end{figure*}

Fig.~\ref{fig3} compares the near-$E_F$ electronic structure of URu$_{2-x}$Fe$_{x}$Si$_{2}$ 
in the HO ($x = 0$) and deep into the AFM 
($x = 0.2$)~\cite{Kanchanavatee2011, Das2015, Ran2016, Butch2016, Williams2017_2} phases,
around the $\textmd{Z}$ and $\Lambda$ high-symmetry points of the BCT Brillouin zone. 
We observe small but unequivocal changes around these points within the first few meV 
below the Fermi level. 
In particular, both in the HO phase, Figs.~\ref{fig3}(a-d), 
and in the AFM phase, Figs.~\ref{fig3}(g-j),
the band dispersions around $\textmd{Z}$ and $\Lambda$
exhibit a clear M-shaped feature with an electron pocket centered at $k_{<100>}=0$. 
However, the band minimum of this electron pocket, 
located at $E-E_F = (-5 \pm 0.5)$~meV in the HO state, 
is pushed up in energy to $E-E_F = (-3 \pm 0.5)$~meV in the AFM state.
This is also seen from the grey momentum-integrated energy distribution curves (EDCs)
around $\textmd{Z}$ and $\Lambda$, shown in Figs.~\ref{fig3}(e,~f) and Figs.~\ref{fig3}(k,~l).
More important, as demonstrated by the momentum-integrated magenta EDCs 
in Figs.~\ref{fig3}(e,~f) and Figs.~\ref{fig3}(k,~l), 
such changes do not correspond to a rigid energy shift, as there are no observable differences 
in the binding energy of the heavy bands at large momenta ($k_{<100>}=0.3$~\AA$^{-1}$), 
far from the hybridization region (around $k_{<100>}=0$).
The Supplementary Information (SI.\ref{SI:Toy-model})
presents complementary analyses of the band structure changes between the HO and AFM phases
around $\textmd{Z}$ and $\Lambda$, as well as a comparison
of the out-of-plane dispersions in both phases (SI.\ref{SI:ARPES-hv-dependence}), 
and a discussion of the changes in electronic structure
at fixed doping ($x=0.2$) across the PM/AFM transition (SI.\ref{SI:PM-AFM})
The latter are qualitatively the same as those observed across
the PM/HO transition~\cite{Bareille2014}, namely the gapping of a large diamond-like
Fermi surface around $\Gamma$ along $k_{<110>}$ and the concomitant
formation of the four Fermi petals, as discussed in Fig.~\ref{fig2}.

The direct observation of a slight but distinct electronic band structure 
change across the HO/AFM phase boundary gives crucial insight to previous results 
obtained by other experimental probes.
Specifically, extremal Fermi surface contours by means of SdH oscillations 
showed only minor differences between the HO and the AFM phases~\cite{Hassinger2010}, 
while transport measurements concluded 
that the partially gapped Fermi surface of the HO phase~\cite{Maple1986} would be further gapped, 
and/or its volume reduced (decrease in the Sommerfeld coefficient),
when the system is driven into the AFM phase~\cite{Jeffries2007, Kanchanavatee2011}. 
Moreover, inelastic neutron scattering experiments suggested that the Fermi surface pockets 
at the $\Gamma$, $\textrm{Z}$ and/or $\Sigma$ (connecting two Fermi petals) wave-vectors 
should slightly distort upon crossing the HO/AFM phase boundary, 
so as to modify the optimal energy for nesting~\cite{Williams2017_2}. 
In agreement with those previous findings 
indicating only subtle Fermi surface changes 
that might be a consequence of better nesting conditions,
our results in the AFM phase, compared to those in the HO phase, 
show a slightly modified electronic structure.
Specifically, we observe an upwards energy shift of the electron pockets 
around the $\textrm{Z}$ and $\Lambda$ high-symmetry points.
Given the isotropic shape of the electron-like pocket around $k =0$, Fig.~\ref{fig2}(c), 
an upward energy shift would be translated into
a concomitant reduction of the corresponding Fermi surface contours,
resulting in a smaller number of charge carriers in the AFM state.
It is worth noting that, according to recent electronic structure calculations,
a decrease in the volume of the unit cell, such as the one induced by chemical pressure
through the partial substitution of Ru with Fe, would induce an energy shift in the opposite
direction to the one experimentally observed 
at the $\textrm{Z}$ and $\Lambda$ points~\cite{Oppeneer2010}. 
We conclude that across the HO/AFM phase boundary, changes 
of the interaction strength between states near the Fermi level
are at the origin of a minor but \emph{essential} 
Fermi surface change in the system, stabilizing one or the other phase.
We thus hope that our work will motivate further theoretical studies 
of the AFM phase of URu$_2$Si$_2$ aiming to couple the observed energy shifts 
with the effect of the AFM nesting vectors~\cite{Oppeneer2011}. 

One may speculate, as predicted by some models~\cite{Dubi2011}, 
that pressure (physical or chemical) could have the effect of decreasing
the Ru~$4d$~-~U~$5f$ interaction strength to a critical value below which the conduction $d$ carriers 
that were coupled to localized $f$ electrons are no longer able to screen 
the magnetic interactions of the latter.
Therefore, at high pressures, due to unscreened magnetic moments, 
the system is driven into the AFM phase.

More generally, within the Mott-Doniach picture 
of heavy-fermion systems~\cite{Mott1974,Doniach1977},
the competition between antiferromagnetic interactions and the Kondo effect
leads to a quantum phase transition between an antiferromagnetic ground state,
characterized by localized-like $f$ electrons, 
and a Kondo lattice of itinerant $f$ electrons.
Our results show that
the interaction strength of near-$E_F$ electronic states
is another crucial ingredient that needs to be taken into account in the understanding 
of the phase diagram and quantum phase transitions of heavy-fermion systems.

\acknowledgments{
We thank Marie-Aude~M\'easson, Sergio G. Magalh\~aes and Peter Riseborough for comments
and discussions.
ARPES work at ISMO was supported by public grants from the French National Research Agency (ANR), 
project Fermi-NESt No ANR-16-CE92-0018.
Single-crystal growth at UCSD was supported by the US Department of Energy, 
Office of Basic Energy Sciences, Division of Materials Sciences and Engineering, 
under Grant No. DE-FG02-04-ER46105. 
Sample characterization at UCSD was sponsored by the National Science Foundation 
under Grant No. DMR-1810310.
}

\subsection*{Data Availability}
All relevant data are contained in this manuscript and the Supplementary Information. 
Reproduction of these data on different samples is available upon reasonable request 
by emailing the corresponding authors.

\subsection*{Author Contributions}
Author contributions:
A.F.S.-S. and M.B.M. designed research;
E.F, J.D., T.C.R., M.G., C.B. and F.F. performed ARPES measurements, 
under the supervision of A.F.S.-S;
E.D.L.R. provided support with instrumentation at UE112-PGM-2b-1$^{3}$ endstation;
S.R., N.K., K.H., N.P., C.T.W., M.B.M. and P.L. prepared and characterized single crystals;
E.F. and C.B., and A.F.S.-S. analyzed and interpreted data;  
E.F. and A.F.S.-S wrote the paper.
All authors discussed extensively the results and the manuscript.

\section*{SUPPLEMENTARY INFORMATION}
\subsection{\label{SI:URS-Xtals} Preparation and characterization of 
UR\lowercase{u}$_{2-x}$F\lowercase{e}$_{x}$S\lowercase{i}$_{2}$ single crystals}
The single crystals of Fe-substituted URu$_2$Si$_2$ on which the ARPES measurements were made 
were grown by the Czochralski method in a tetra-arc furnace at UCSD. 
These crystals were characterized at UCSD by means of X-ray diffraction, electrical resistivity, 
magnetization, specific heat, and thermal expansion measurements as a function of temperature, 
magnetic field and pressure as described in several publications~\cite{Ran2016,Wolowiec2016,Ran2017}.  
The electrical resistivity measurements were performed using a home-built probe 
in a liquid $4$He Dewar by means of a standard four-wire technique at $16$~Hz, 
using a Linear Research LR700 a.c. resistance bridge. 
Magnetization measurements were made in a magnetic field of $0.1$~T, 
using a Quantum Design magnetic property measurement system (MPMS). 
Specific heat measurements were performed in a Quantum Design DynaCool 
physical property measurement system (DC-PPMS-9), using a heat-pulse technique. 
Thermal expansion measurements were made in a Quantum Design DC-PPMS-9 
with a dilatometer measurement option (model P680).  
The single crystals with $x = 0$ (pure URu$_2$Si$_2$) and $x > 0$ prepared at UCSD 
were further characterized by means of spectroscopic measurements, 
performed in the laboratories of collaborators or national laboratories, 
which include neutron diffraction~\cite{Das2013,Das2015}, 
inelastic neutron scattering~\cite{Butch2015,Butch2016}, 
ultrasonic~\cite{Yanigasawa2013,Yanigasawa2018}, infrared spectroscopy~\cite{Hall2015}, 
Raman spectroscopy~\cite{Kung2016}, ultrafast optical spectroscopy~\cite{Kissin2019}, 
X-ray scattering~\cite{Wray2015,Amorese2020}, EXAFS~\cite{Bridges2020}, 
and quasiparticle scattering~\cite{Zhang2020} measurements.    
A simple measurement that is often used to assess the quality of a metallic sample 
is the electrical resistivity.  The electrical resistivity $\rho (T)$ measurements 
in the vicinity of the transition temperature for the single crystal samples 
of URu$_{1-x}$Fe$_x$Si$_2$ with $x = 0$,~$0.1$ and~$0.2$ are shown in Refs.~\cite{Ran2016} 
and~\cite{Wolowiec2016}. The width of the resistive transitions for the samples 
in the LMAFM phase with Fe concentrations of $x = 0.1$ and~$0.2$ are only slightly broadened 
when compared to those reported for the pristine URu$_2$Si$_2$ sample in the HO phase. 
The sharp transitions in the single crystal samples of URu$_{1-x}$Fe$_x$Si$_2$ 
with $x = 0.1$ and~$0.2$ indicate the samples are of high quality. 
However, there is a certain amount of broadening of the transition, 
which is to be expected with an increase in Fe concentration.  
Here, the larger concentrations of Fe that are introduced into the melts grown 
by the Czochralski method can result in a larger degree of inhomogeneity within the sample, 
especially in the direction of the vertical axis of the rod-shaped boule 
that is pulled from the melt.  Furthermore, the transition temperatures 
of $T_0 \approx 17$,~$18$,~and~$21$K, for the $x = 0$,~$0.1$,~and~$0.2$ samples, 
respectively, are consistent with the behavior of the phase boundary 
in both the HO and LMAFM phases across a broad range of Fe concentrations, e.g., $T_0$ vs $x$ 
in Ref.~\cite{Ran2016} and $T_0$ vs $x, P$ in Ref.~\cite{Wolowiec2016}.

\subsection{\label{SI:ARPES-details} ARPES measurements}
ARPES measurements were performed at the UE112-PGM-2b-1$^{3}$ endstation of BESSY II 
with a Scienta R4000 hemispherical electron analyser. 
Data for angle-resolved measurements were collected with photon energies between 
20 and 67~eV. Higher photon energies were used for angle-integrated measurements,
(see SI.\ref{SI:ARPES-SurfaceTerm}).
The instrumental resolution varied from 3 to 7~meV according to experimental conditions.
The light polarisation was linear horizontal or linear vertical as mentioned in the figure captions. 
The size of the beam spot was 40~$\mu$m. Samples were cleaved in situ at a temperature below 20K. 
The pressure was at all times lower than $1.0 \times 10^{-10}$~mbar. 
The raw ARPES spectra were normalized to the total intensity of the energy distribution curves 
and -if necessary- served to calculate the 2D curvature~\cite{Zhang2011}. 
All data comparisons refer to identical experimental conditions and data treatment.
The results have been reproduced in at least five different cleaves.
The location of the normal-emission high-symmetry $\Gamma$, $\Lambda$ and Z points
was always double-checked by performing in advance in-plane Fermi-surface maps 
in steps of $0.5$ or $0.25$ degrees around these points.

Similar to previous ARPES studies~\cite{Boariu2013, Bareille2014}, 
we note that although our low-temperature data (1K) were acquired at 
slightly below the superconducting (SC) transition for $x = 0$, 
the corresponding energy gap~\cite{Hasselbach1992} is at least 
1-2 orders of magnitude smaller than our experimental resolution, 
and any possible changes associated with superconductivity cannot be seen in the present study.   

\subsection{\label{SI:3D-mapping} 3D $k$-space mapping}
Within the free-electron final state model, ARPES measurements at constant photon energy 
give the electronic structure at the surface of a spherical cap of radius
$k = \sqrt{2m_\text{e}/\hbar^2} \left(h\nu - \Phi + V_0 \right)^{1/2}$. 
Here, $m_\text{e}$ is the free electron mass, $\Phi$ is the work function, 
and $V_0 = 13$~eV is the ``inner potential" 
of URu$_2$Si$_2$~\cite{Denlinger2001,Santander2009,Boariu2010,Bareille2014}.
Measurements around normal emission provide the electronic structure in a plane 
nearly parallel to the surface plane.
Likewise, measurements as a function of photon energy provide the electronic structure in a plane
perpendicular to the surface.

\subsection{\label{SI:ARPES-SurfaceTerm} Electronic structure of 
(F\lowercase{e}-doped) UR\lowercase{u}$_{2}$S\lowercase{i}$_{2}$ 
as a function of surface termination}
\begin{figure*}[!p]
	\centering
		\includegraphics[clip, width=\textwidth]{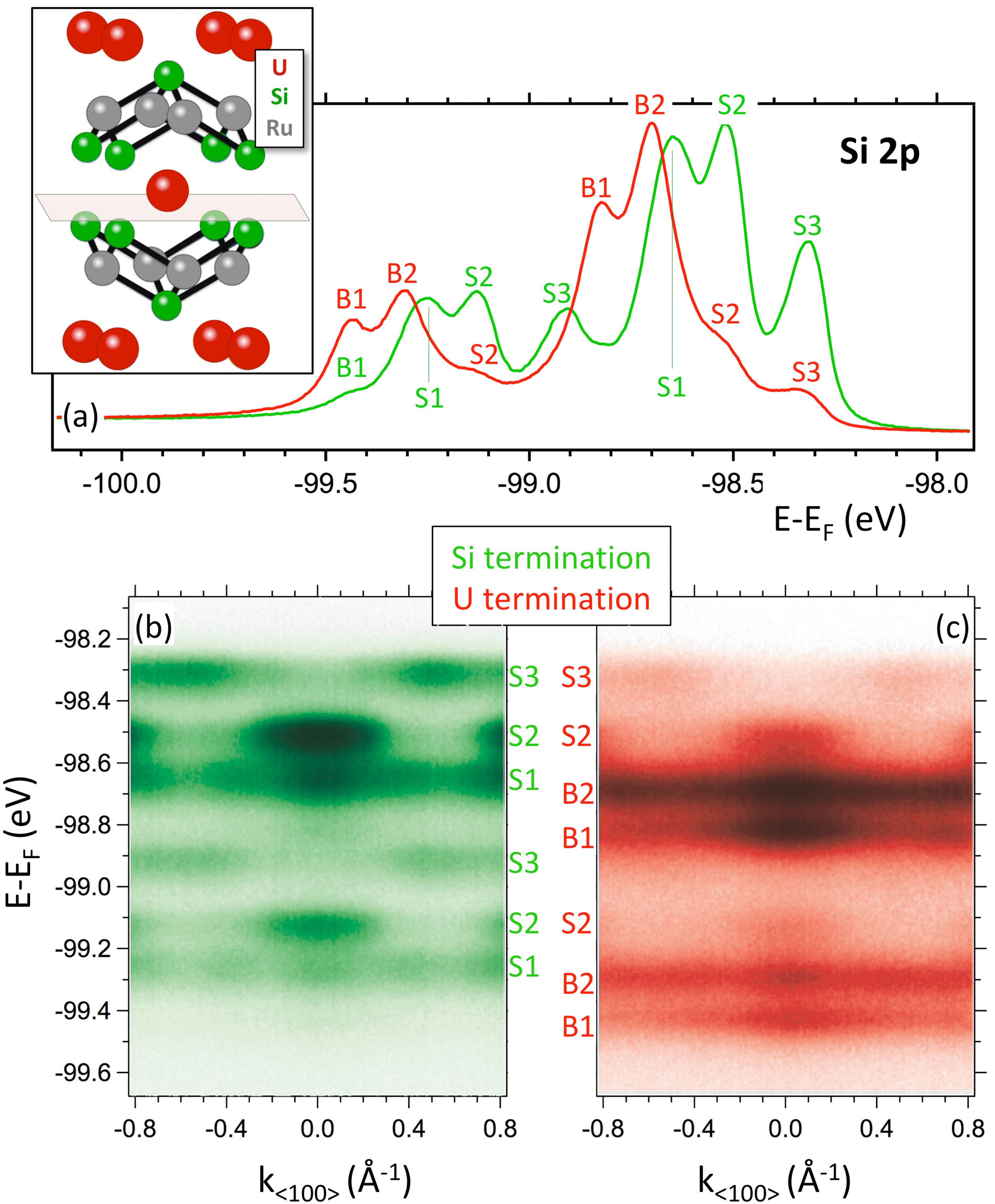}
    	\caption{
        	 (a) Angle-integrated photoemission spectra of the Si~$2p$ core level peaks 
        	 from a sample location with a dominant Si (green) or U (red) surface termination.
        	 `B' and `S' denote the bulk and surface origin of the corresponding features. 
        	 The inset is a sketch of the unit cell of the parent compound 
        	 and of a typical cleavage plane. 
        	 (b,~c) ARPES spectra of the Si~$2p$ core level peaks shown in (a). 
        	 Results on a Si-terminated (U-terminated) surface are represented 
        	 by green (red) spectral lines and green- (red-) hued images. 
        	 Data have been acquired at 1 K using 50.5~eV photons ($3^{\textmd{rd}}$-order harmonics) 
        	 with linear vertical polarisation.
            }
	\label{figS1}
\end{figure*}

\begin{figure*}[!p]
	\centering
        \includegraphics[clip, width=\textwidth]{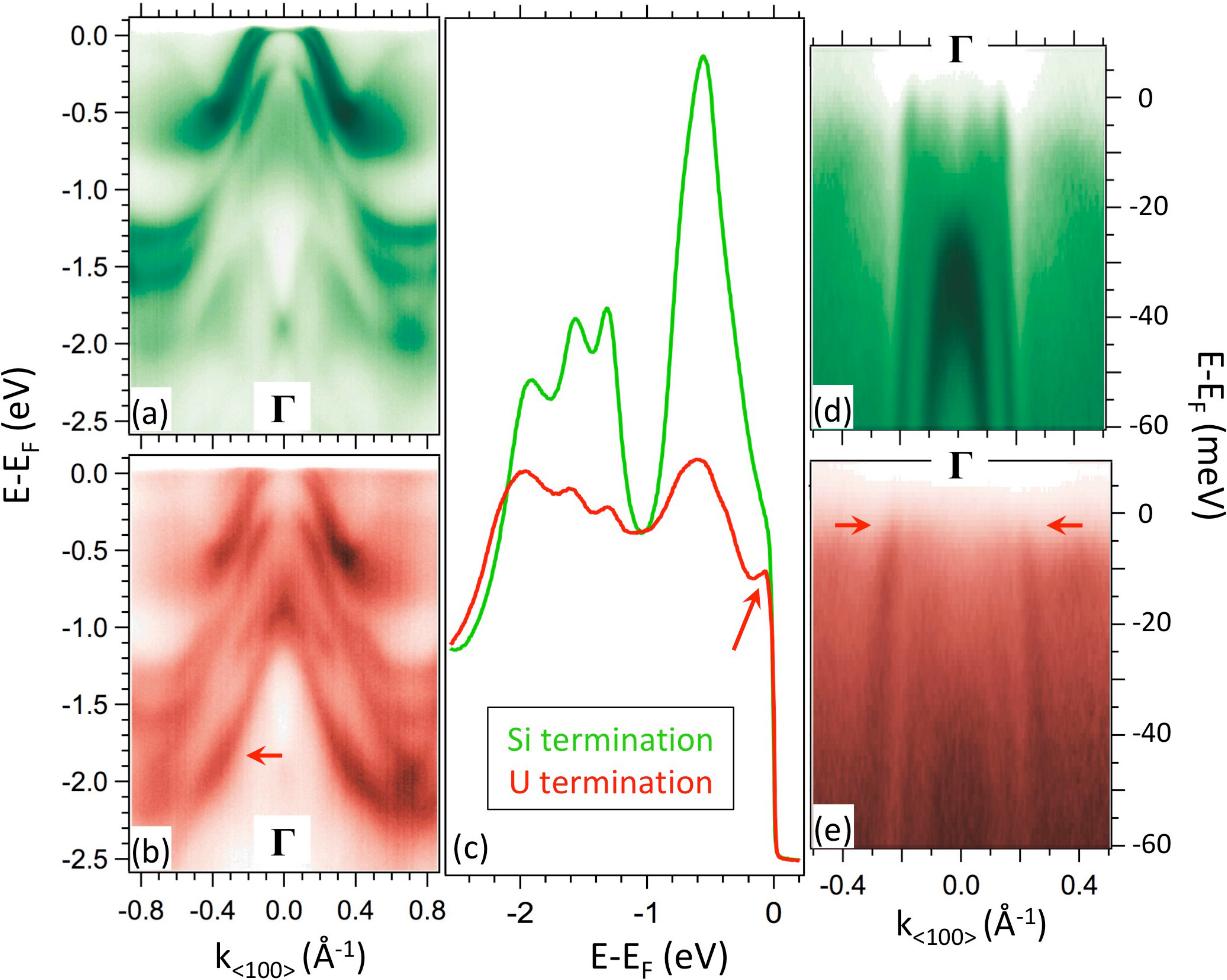}
    	\caption{
        	 (a,~b) Energy dispersion of the valence band in a Si-terminated (top) 
        	 and U-terminated (bottom) surface. The energy dispersions are acquired 
        	 in sample locations that exhibit the Si 2$p$ core level profiles of Fig.~\ref{figS1}. 
        	 The arrow in (b) points at a state that is only observed 
        	 at the U-terminated surface. 
        	 (c) Angle-integrated photoemission spectra of the valence band 
        	 in both terminations (green~=~Si-termination, red~=~U-termination). 
        	 The arrow points at a peak of presumably U 5$f$ origin that corresponds
        	  to a streak of $k$-independent intensity. 
        	  (d,~e) A zoom in the near-$E_F$ electronic structure reveals different bands 
        	  crossing the fermi level for each termination. 
        	  The arrows in (e) point at a state that is only observed at the U-terminated surface. 
        	  Results on a Si-terminated (U-terminated) surface are represented 
        	  by green (red) spectral lines and green- (red-)hued images. 
        	  Data have been acquired at 1~K using 50.5~eV photons with linear vertical polarisation.
        }
	\label{figS2}
\end{figure*}

\begin{figure*}[!p]
	\centering
        \includegraphics[clip, width=\textwidth]{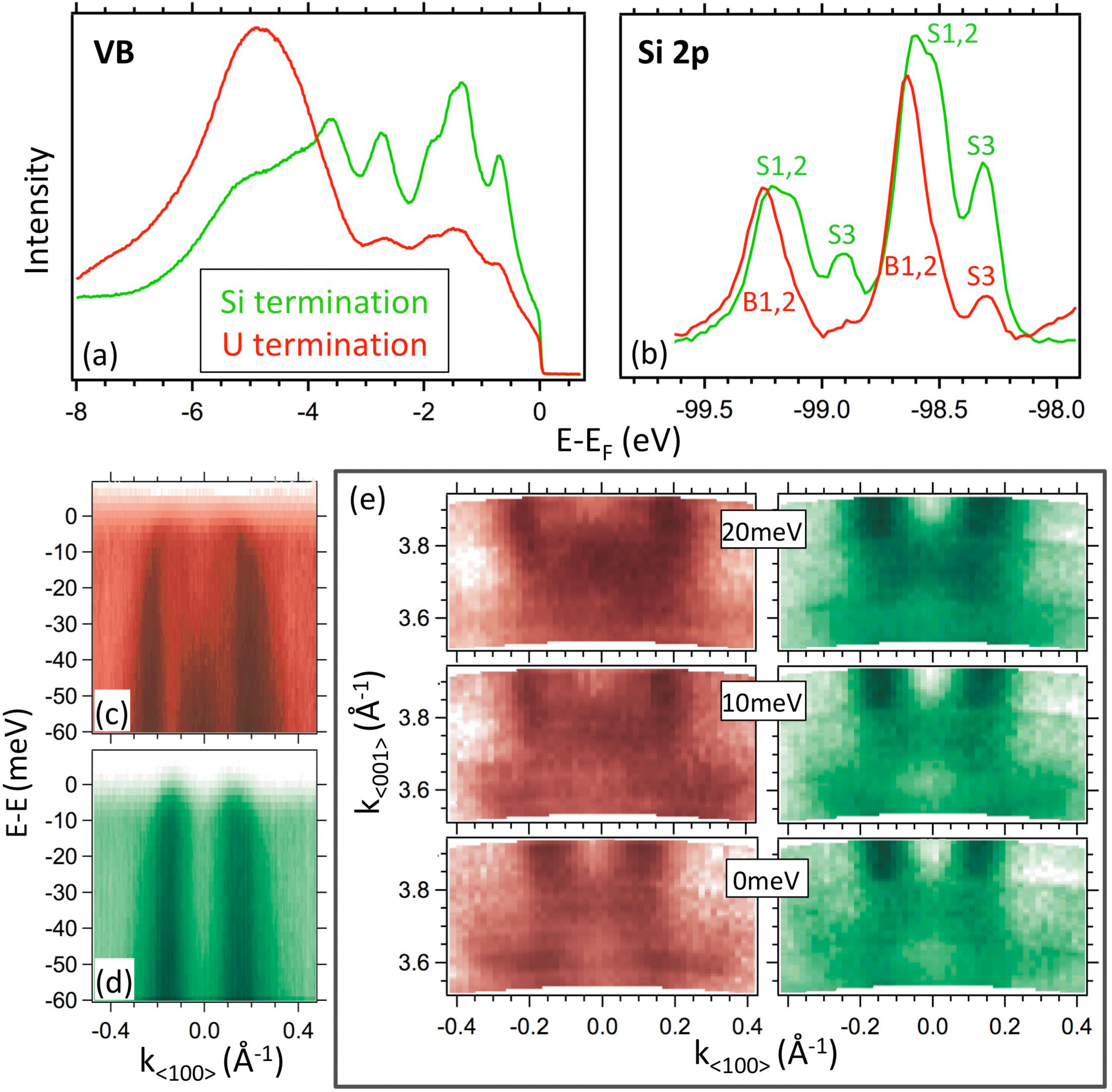}
    	\caption{
        	 (a,~b) Angle-integrated photoemission spectra of the valence band 
        	 and the Si~$2p$ core level peaks in a Si-terminated (green) 
        	 and U-terminated (red) surface. `B' and `S' denote the bulk or surface origin 
        	 of the corresponding features. 
        	 (c,~d) Energy dispersion of the electronic states near $E_F$. 
        	 (e) Constant energy maps on a plane perpendicular to the sample surface. 
        	 Data in panels (a) and (b) were acquired at 70~K using 90~eV photons
        	 ($2^{\textmd{nd}}$-order harmonics for Si 2$p$) with linear horizontal polarisation. 
        	 Data in panels (c-e) were acquired at 1~K using linear vertical polarisation.
        	 The energy dispersions were obtained with 49~eV photons 
        	 and the constant energy maps by varying the photon energy between 39 and 51~eV. 
        	 Green- (red-) hued images correspond to the Si- (U-) terminated surface.
      }
	\label{figS3}
\end{figure*}

The crystal structure of URu$_{2}$Si$_{2}$ is shown in the inset of Fig.~\ref{figS1}(a).
The lattice has a body-centred tetragonal symmetry with Ru, Si and U atoms 
forming in-plane square lattices and each Ru tetrahedrally coordinated 
to four Si atoms~\cite{Ban1965}. In Fe-doped URS (Fe-URS), 
there is a partial substitution of Ru by Fe.
Due to the smaller size of the Fe atoms, a reduction of the unit cell volume may be expected, 
producing an effective chemical pressure~\cite{Kanchanavatee2011}. 
The characteristic -U-Si-Ru-Si-U- atomic stacking permits fracturing in UHV conditions
to expose (001) cleavage surfaces. A cleavage plane is indicated 
in the inset of Fig.~\ref{figS1}(a) and may yield a Si- or a U-terminated surface. 

Despite important advances in revealing the electronic band structure of URS 
by means of ARPES~\cite{Denlinger2001, Santander2009, Yoshida2010, Kawasaki2011, 
Yoshida2012, Yoshida2013, Meng2013, Boariu2013, Chatterjee2013, Bareille2014, Fujimori2016} 
and in visualising different terminations in the nanometer scale by means of STM~\cite{Aynajian2010}, 
there are scarce reports including an effort to combine these findings 
\footnote{The electronic band structure of different terminations by photoemission 
has been the focus of studies of isostructural compounds such as CeRu$_{2}$Si$_{2}$ 
and YbRh$_{2}$Si$_{2}$ \cite{Denlinger2001, Denlinger2002, Dazenbacher2007, Fujimori2016}.}. 
Unveiling the relationship between surface termination and electronic band structure 
is of fundamental importance as surface effects can dominate the ARPES spectra 
and thus complicate the interpretation of electronic band structure changes 
when comparing samples from different regions of the phase diagram. 

Fig.~\ref{figS1}(a) presents the Si~$2p$ core level spectra from different areas 
of a Fe-URS (001) cleavage surface. Si surface sites are expected to give rise 
to satellites of the main Si~$2p$ bulk core levels peaks. 
We therefore attribute the green (red) curve to a sample location where the surface termination 
is dominated by Si (U) atoms. `B' and `S' denote the bulk and surface origin of the corresponding peaks, 
while the numbers group together the spin-orbit counterparts of each Si site. 
The red curve reveals that there are more than one Si bulk sites, 
in agreement with the URu$_{2}$Si$_{2}$ crystal structure. 
On comparison of the two curves, we note that there are no energy shifts, 
but only changes in the relative intensity of the peaks. 

Fig.~\ref{figS1}(b,~c) prove that each core level peak exhibits 
a characteristic intensity distribution in $k$-space, due to photoelectron diffraction. 
Since the intensity distribution is expected to vary only when the associated Si site becomes different, 
this observation permits us to unequivocally associate each peak to its spin-orbit counterpart 
and to the corresponding peak of the other termination. 
Taking advantage of their intensity distributions, one can be certain that the features 
termed S1 and S2 correspond to different Si sites from those termed B1 and B2. 
As expected, the spectroscopic fingerprint of S1, S2 and S3 is very weak on the surface 
with a termination layer dominated by U atoms. A well-defined photoelectron diffraction pattern 
from surface sites proves that the surface atoms are arranged into a periodic lattice 
rather than in a number of random steps and edges. Therefore, our results suggest 
that areas of a \textit{dominant} Si- (or U-) termination layer can be at least as large 
as our beam spot (40 $\mu$m), although terraces with a strictly unique atomic termination 
are much smaller~\cite{Aynajian2010}. 

The valence band spectra in Figs.~\ref{figS2}(a,~b) show that dispersing states are present 
in both terminations. The U-terminated surface 
exhibits additional non-dispersing spectral weight just below the Fermi level, 
which is presumably of U 5$f$ origin [arrow in Fig.~\ref{figS2}(c)], but also a strongly dispersing state 
within the energy range of 1-2 eV that is centred at $\Gamma$ and has no counterpart 
in the Si-terminated surface [arrow in Fig.~\ref{figS2}(b)]. 
Moreover, looking closer at the electronic structure within the first few meV below $E_{F}$, 
the U-terminated surface presents a rapidly dispersing hole-like state crossing the Fermi level 
[arrows in Fig.~\ref{figS2}(e)] that cannot be identified with any of the states 
observed on the Si-terminated surface [Fig.~1 of the main text and Fig.~\ref{figS2}(d)]. 
The presence of dispersing states that are unique to each surface termination 
confirms that both surfaces are well-ordered.
Fig.~\ref{figS3} summarizes our results on URu$_{1.8}$Fe$_{0.2}$Si$_{2}$. 
Despite changes in stoichiometry and differences in experimental conditions 
(e.g. photon energy, energy resolution) between Figs.~\ref{figS1}/\ref{figS2} and Fig.~\ref{figS3},
one can draw common conclusions for the two terminations. 
Most dispersing states within the valence band continuum are common to both terminations, 
but there are certain electronic states that are termination-specific.
For the Si-terminated surface, these are a hole-like parabolic surface state 
and the light-hole conduction band --see Fig.~\ref{figS3}(d).
On the other hand, the fingerprints of the the U-terminated surface
are a strongly dispersing band with a maximum at 1 eV below $E_{F}$ [Fig.~\ref{figS2}(b)]
and the hole-like electronic state crossing $E_{F}$ [Figs.~\ref{figS2}(e) and \ref{figS3}(c)].
The angle-integrated spectrum of the valence band in the U-termination is nevertheless
dominated by a broad peak at a binding energy of 5~eV --see Fig.~\ref{figS3}(a).
In Fig.~\ref{figS3}(e), we have followed the out-of-plane dispersion of the low energy states 
that are characteristic for each termination. Although there are no clear differences at $E_{F}$, 
data at slightly higher binding energies show that the contours in the U-terminated surface 
are found at larger $k$ values than those in the Si-terminated surface. 
This observation is agreement with a comparison of Fig.~\ref{figS3}(c) and \ref{figS3}(d) where
the light-hole conduction band is somewhat wider in momentum in the U termination.
Last but not least, there are important differences in the Si 2$p$ peaks, 
which have been already addressed in the discussion of Fig.~\ref{figS1}: 
the surface-related (bulk-related) Si peaks are enhanced in the Si-terminated (U-terminated) surface. 
As shown in Fig.~\ref{figS3}(b), we note that the Si 2$p$ core level peaks for the two terminations 
can be perfectly reproduced under different experimental conditions.

\subsection{\label{SI:Robustness-cleave-surface} Robustness of the 
UR\lowercase{u}$_{2-x}$F\lowercase{e}$_{x}$S\lowercase{i}$_{2}$ 
cleavage surfaces in a UHV environment}
\begin{figure*}[!t]
	\centering
        \includegraphics[clip, width=\textwidth]{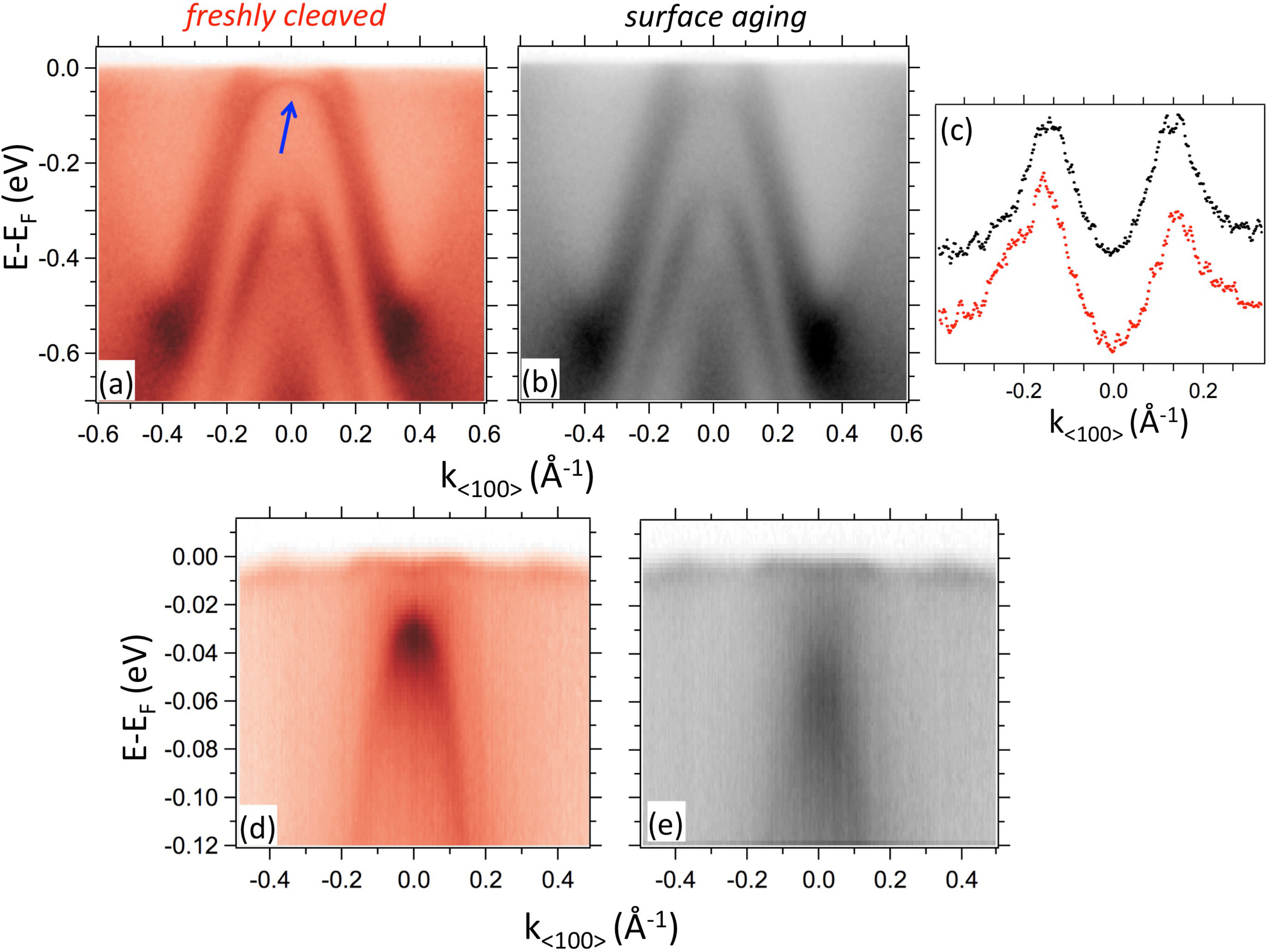}
    	\caption{
        	 (a,~b) Valence band of URu$_{2}$Si$_{2}$ measured 1 hour (red hues) 
        	 and 26 hours (grays) after cleavage. 
        	 The arrow shows the surface state.
        	 (c) Momentum distribution curves at $E_F$ (integration $\pm$10~meV) 
        	 of the ARPES intensity maps shown in panels (a) and (b). 
        	 (d,~e) Near-$E_F$ electronic dispersion measured 3 hours (red hues) 
        	 and 26 hours (grays) after cleavage. 
        	 Comparisons are performed on the same cleavage surface and each gray-scale 
        	 data set includes a 12-hour period of non-exposure to photons. 
        	 Data in panels (a-c) have been acquired using 50.5~eV photons with LV polarisation. 
        	 Data in panels (d,~e) have been acquired using 24~eV photons with LV polarisation.
        	 The temperature was 1K and the pressure lower than $1.0 \times 10^{-10}$~mbar.
        }
	\label{figS4}
\end{figure*}

\begin{figure*}[!t]
	\centering
        \includegraphics[clip, width=0.95\textwidth]{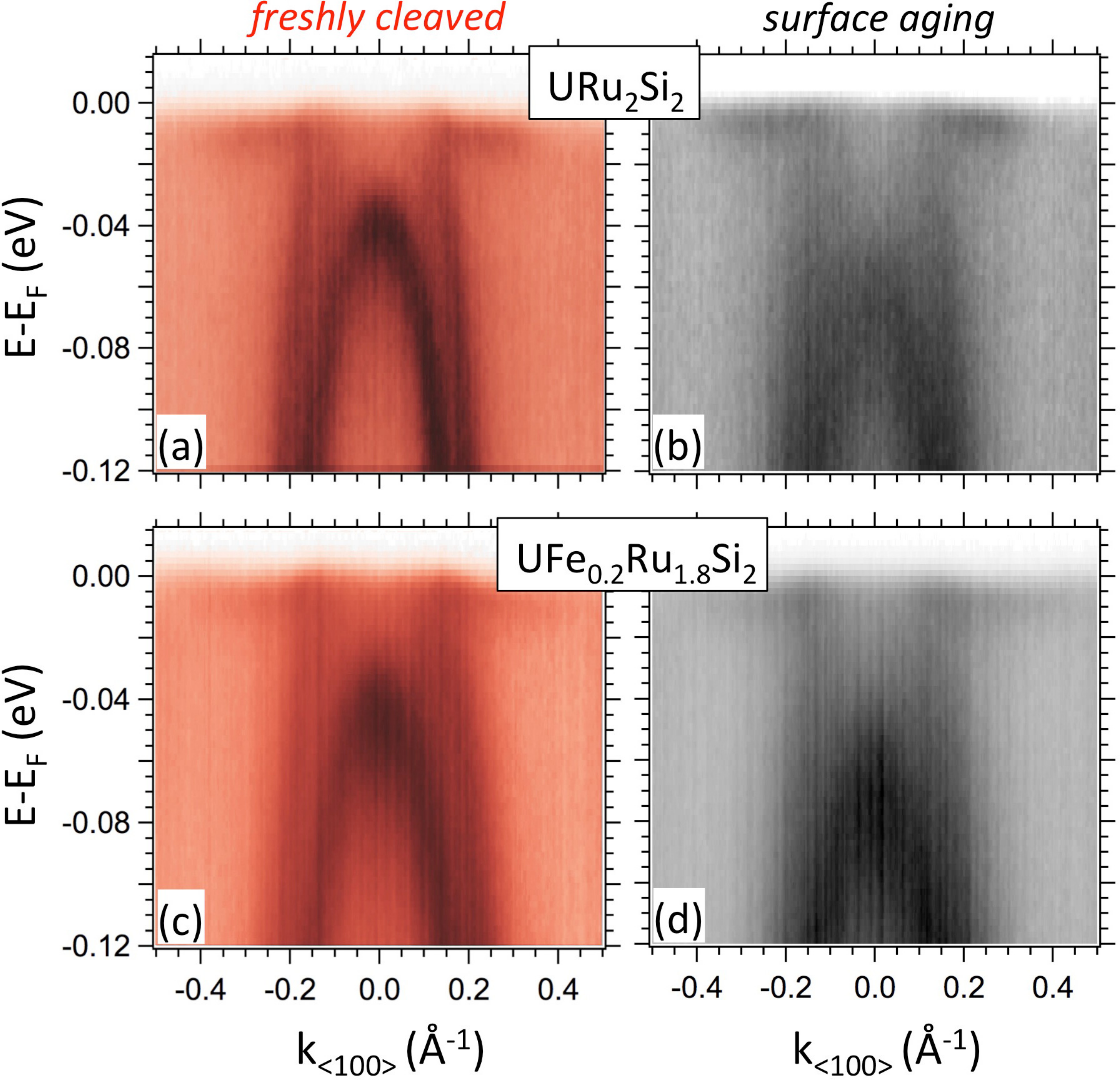}
    	\caption{
        	 (a,~b) Near-$E_F$ electronic dispersion of URu$_{2}$Si$_{2}$ 
        	 measured 6 hours (red hues) and 26 hours (grays) after cleavage. 
        	 There is a 12-hour period of non-exposure to photons between panels (a) and (b). 
        	 (c,~d) Near-$E_{F}$ electronic dispersion of URu$_{1.8}$Fe$_{0.2}$Si$_{2}$ 
        	 measured 4 hours (red hues) and 16 hours (grays) after cleavage. 
        	 No exposure to photons has been performed between panels (c) and (d). 
        	 Data in all panels have been acquired using 50.5~eV photons with LV polarisation. 
        	 The temperature was 1K and the pressure lower than $1.0 \times 10^{-10}$~mbar.
        }
	\label{figS5}
\end{figure*}

The Si-terminated cleavage surfaces of URS and Fe-URS are sensitive 
to the residual gases in UHV conditions. 
Figs.~\ref{figS4} and \ref{figS5} present a comparison of the ARPES data acquired 
on a freshly cleaved (Fe)-URu$_{2}$Si$_{2}$ surface (red-hued images) 
and on the same surface after intentional exposure to residual UHV gases for a few hours 
(gray-scale images). The hole-like surface state is dramatically affected by aging: 
its spectral signature becomes broader, while its maximum shifts to higher binding energy.
It is precisely the sensitivity of this spectral feature to surface contamination
that demonstrated its surface origin in previous works~\cite{Boariu2010}. 
The degradation rate and the size of the energy shift 
depend on the residual gas pressure and the presence of excited molecular species~\cite{Boariu2010}. 
As seen in Figs.~\ref{figS4}(d,~e), a base pressure of $1.0 \times 10^{-10}$~mbar 
induces a shift of the order of 25~meV in 23~hours. This rate is smaller by a factor of 10-20 
with respect to the results presented in previous works~\cite{Boariu2010}, 
a study performed under similar UHV conditions but using a He discharge lamp, 
whose plasma can produce a considerable amount of excited species coming from contaminants in the gas. 
We note that the energy shift and degradation of the surface state is appreciable 
even in the absence of incoming photons [Fig.~\ref{figS5}, bottom]. 
We therefore exclude photo-ionisation and photo-dissociation of residual gas molecules 
as mechanisms of surface degradation. 

The rest of the electronic structure is also affected by surface aging, 
albeit to a smaller extent than the surface state. The light-hole conduction band presents 
a decreased signal-to-noise ratio after exposure to residual gas molecules (Fig.~\ref{figS5}).
Despite the degradation, there is no observed energy shift of the light-hole conduction band, 
as its Fermi wave-vectors do not change before and after exposure 
to residual gases [Fig.~\ref{figS4}(c)]. It is noteworthy that the heavy bands just below $E_{F}$
are the least affected by surface contamination [Fig.~\ref{figS4}(bottom) and Fig.~\ref{figS5}].
Variations in the sensitivity of different bulk states to surface aging 
can be an additional proof of their orbital origin. 
In a Si-terminated surface, as shown in Fig.~\ref{figS1}(a), U atoms are far from the topmost layer,
thus states of U 5$f$ origin would be indeed the least sensitive to surface contamination.

\subsection{\label{SI:Toy-model} Further comparison and simplified modeling 
of the near-$E_F$ electronic structure of pure and F\lowercase{e}-doped
UR\lowercase{u}$_{2}$S\lowercase{i}$_{2}$}
\begin{figure*}[!p]
	\centering
        \includegraphics[clip, width=0.9\textwidth]{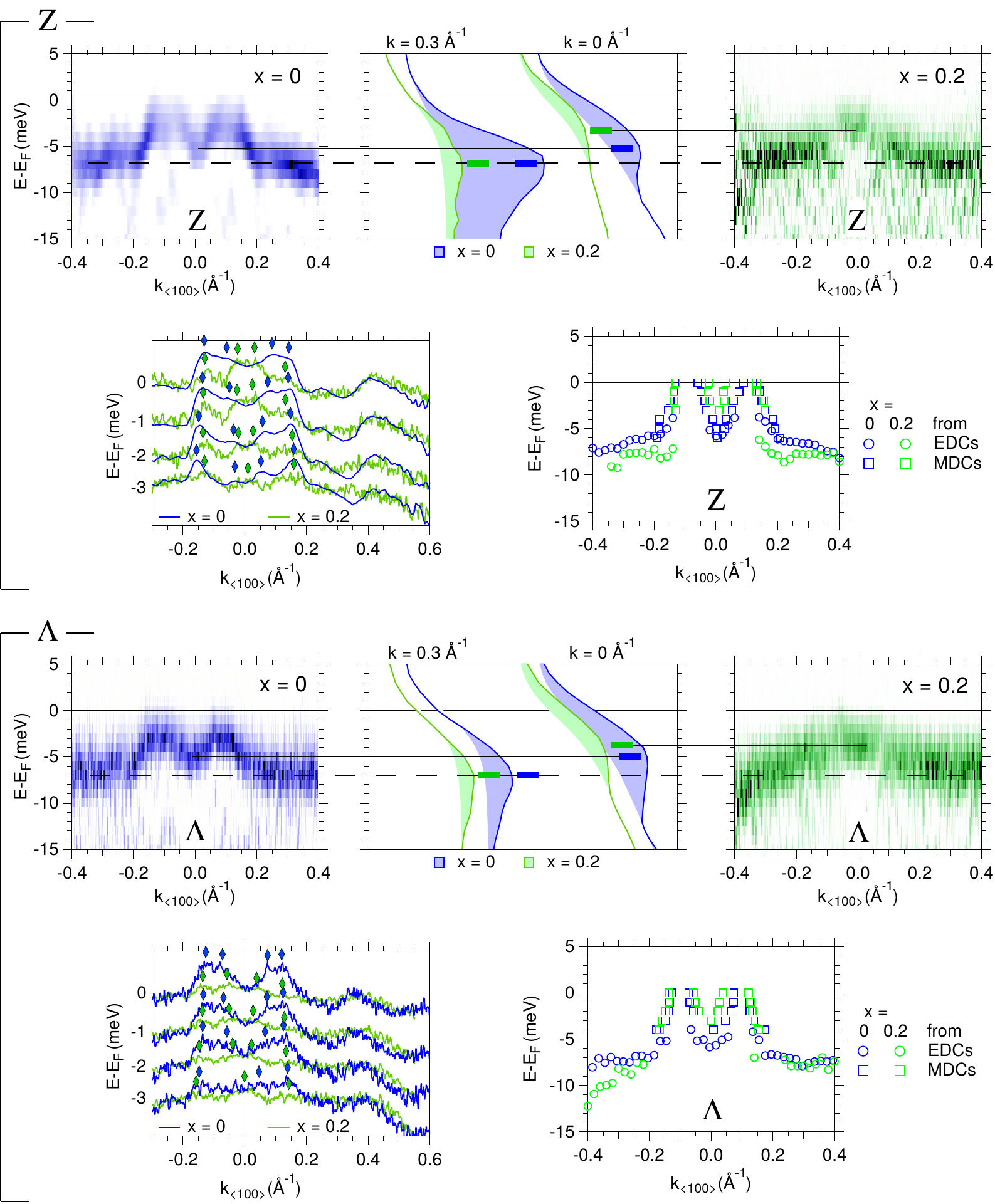}
    	\caption{
        	 (a,~b)~ARPES energy-momentum maps (curvature spectra) along $k_{<100>}$ around 
        	 the Z point in, respectively, pure URu$_{2}$Si$_{2}$ (HO state, blue hues) and 
        	 URu$_{1.8}$Fe$_{0.2}$Si$_{2}$ (AFM state, green hues).
        	 The continuous and dashed horizontal lines show, respectively, 
        	 the bottoms of the electron-pockets at Z and the energy of the heavy tail
        	 of the M-shaped band at higher momenta.
        	 (c)~Corresponding energy distribution curves at $k_{<100>}=0$ (Z point) and at 
        	 $k_{<100>}=0.3$~\AA$^{-1}$. 
        	 The horizontal thick bars mark the peak positions of the different EDCs.        	 
        	 Blue curves and bars, HO state; green curves and bars, AFM state.
        	 (d)~Momentum distribution curves (MDCs) at $E-E_F = 0, -1, -2, -3$~meV.
        	 Diamond markers show the MDCs peaks. 
        	 Blue curves and markers, HO state; green curves and markers, AFM state.
        	 (e)~Band dispersions along $k_{<100>}$ around the Z point in, respectively, 
        	 pure URu$_{2}$Si$_{2}$ (HO state, blue markers) and 
        	 URu$_{1.8}$Fe$_{0.2}$Si$_{2}$ (AFM state, green markers), 
        	 extracted from the EDCs (circles) and MDCs (squares) peak positions.
        	 (f-j)~Data analogous to (a-e) around the $\Lambda$ point in 
        	 pure URu$_{2}$Si$_{2}$ (HO state, blue hues, curves, bars and markers) 
        	 and URu$_{1.8}$Fe$_{0.2}$Si$_{2}$ (AFM state, green curves, bars and markers).
        }
	\label{figS6}
\end{figure*}

\begin{figure*}[!t]
	\centering
        \includegraphics[clip, width=0.9\textwidth]{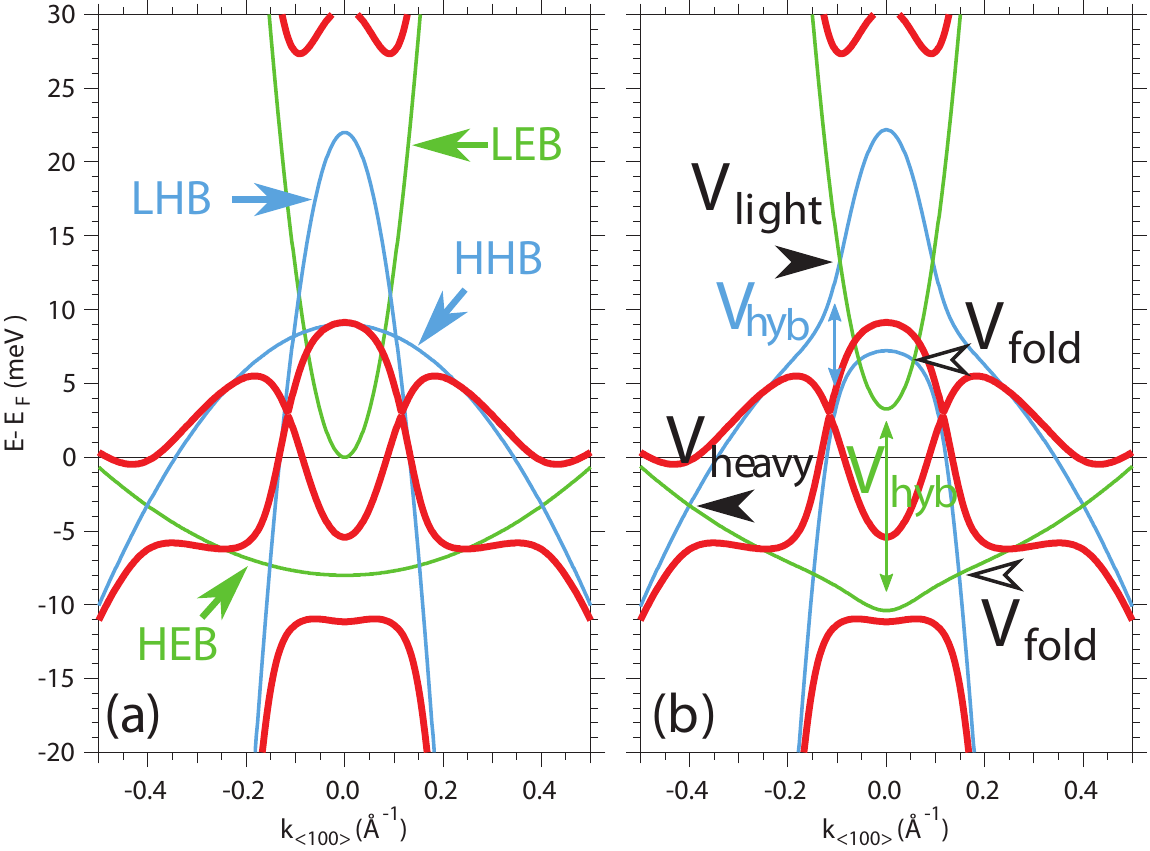}
    	\caption{
        	 (a) The four bands of the phenomenological toy model without interactions 
        	 (blue and green curves) and with interactions that simulate the HO phase (red curves). 
        	 (b) The four bands of the phenomenological toy model either in the presence 
        	 of only the Kondo interaction $\Vhyb$ (blue and green curves), 
        	 or in the presence of all interactions (red curves). 
        	 As in (a), the values of the interactions simulate the HO phase. 
        	 In panel (b), arrows point out the energy gaps due to the different interactions. 
        	 We note that $\Vhyb$ between the LEB and the HEB does not induce 
        	 a new energy gap but slightly modifies the energy position 
        	 and the effective masses of these two bands.
        }
	\label{figS7}
\end{figure*}

\begin{figure*}[!t]
	\centering
        \includegraphics[clip, width=0.9\textwidth]{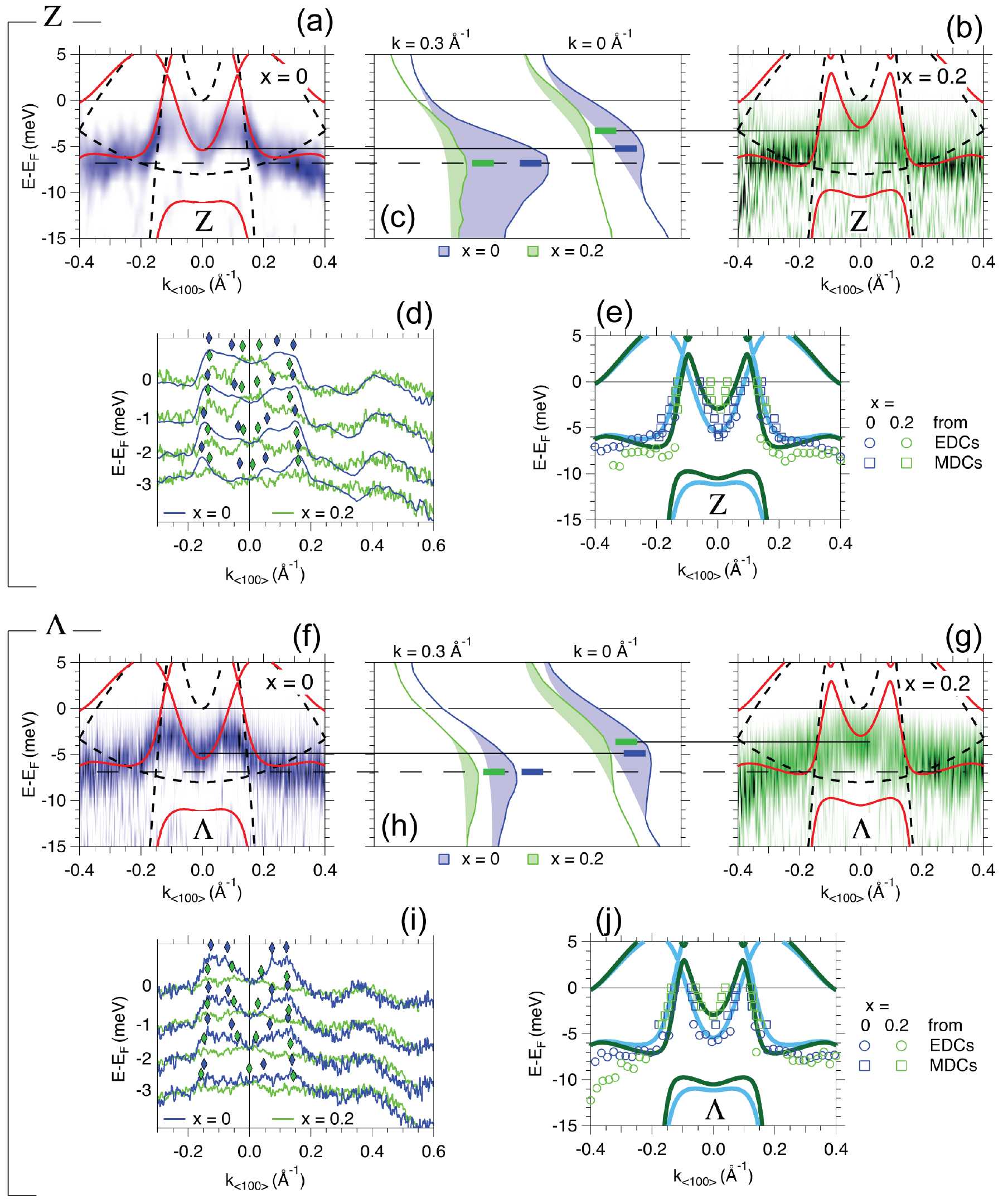}
    	\caption{
        	 Same data as Fig.~\ref{figS6}, with the fits resulting from the toy model 
        	 superimposed.
        	 In panels (a,~b) and (f,~g), the red curves show the dispersion of the toy model bands 
        	 with the interaction parameters specified in Table~\ref{tab:table3}. 
        	 The dashed black curves show the dispersion of the parent parabolas 
        	 in the absence of any interaction.
        	 In panels (e) and (j), the blue and green solid lines show the dispersion 
        	 of the toy model bands, traced with red lines in (a,~b).
        }
	\label{figS8}
\end{figure*}

Fig.~\ref{figS6} presents complementary data showing the decrease in size 
of the electron pockets around Z and $\Lambda$ in the AFM state
(URu$_{1.8}$Fe$_{0.2}$Si$_{2}$ sample, green hues, curves, bars and markers) 
compared to the HO state (URu$_{2}$Si$_{2}$ sample, blue colors).
Figs.~\ref{figS6}(a,~b) show the ARPES energy-momentum dispersion around the Z point in, 
respectively, pure URu$_{2}$Si$_{2}$ and URu$_{1.8}$Fe$_{0.2}$Si$_{2}$.
The corresponding energy distribution curves at $k_{<100>}=0$ (Z point) and at 
$k_{<100>}=0.3$~\AA$^{-1}$ (heavy tail of the M-shaped band) are shown
in Fig.~\ref{figS6}(c). One observes that, while the heavy band at high momenta
is essentially unaffected by the HO/AFM transition, the electron pocket
at Z is shifted up in energy: its band bottom moves up by about $2$~meV.
This is corroborated by the analysis of the MDC and EDC peak positions,
Figs.~\ref{figS6}(d,~e), which show a concomitant reduction 
of the Fermi momenta of the electron pocket in the AFM state
--see in particular the complete band dispersions of the M-shaped bands, 
extracted from the MDC and EDC peaks, in Fig.~\ref{figS6}(e).
As demonstrated by Figs.~\ref{figS6}(f-j), the electron pocket around $\Lambda$
shows a comparable up-shift in energy, of about $1.5-2$~meV, as well as a reduction 
of the Fermi momenta in the AFM state.

The band structure we observe at the vicinity of the Fermi level around normal emission 
can be fairly described with a toy model, 
introduced in our previous works~\cite{Boariu2013, Bareille2014}, 
generalized here to four parabolic ``parent'' bands: one heavy electron-like band (HEB),
one light electron-like band (LEB), one heavy hole-like band (HHB), 
and one light hole-like band (LHB). 
These four parent bands, shown in panel (a) of Fig.~\ref{figS7},
are then subject to pair-wise interactions (see below).
As pointed out in Ref.~\cite{Bareille2014}, the effective bands resulting from our model
compare well (modulo the experimentally observed mass renormalization) 
with DFT calculations near $E_{F}$~\cite{Oppeneer2010}. 
Moreover, they are reminiscent of the results of more recent DFT calculations 
that revealed electron-like contours around $\Gamma$ 
and hole-like contours around $\textmd{Z}$~\cite{Kawasaki2011}.

Given the weak out-of-plane dispersion of the bands around $k_{\|}=0$ 
(Supplementary Information, SI.\ref{SI:ARPES-hv-dependence}),
we used the same toy model to fit the band dispersion around the $\textmd{Z}$ 
and $\Lambda$ high symmetry points, both in the HO and AFM phases. 
The results of the fit, discussed next, 
are superimposed on the experimental data in Fig.~\ref{figS8}.
The fitting masses of the different parent bands were 
$\mleb \approx 3 \me$, $\mlhb \approx -3 \me$, $\mheb \approx 130 \me$, $\mhhb \approx -50 \me$. 
The large difference in effective masses between the light and heavy bands 
suggests that the latter are of predominant U~$5f$ character, while the light bands 
are of predominant Ru~$4d$ character. 
The effective masses and the energy position of the parent bands are summarized 
in Table~\ref{tab:table1}.

\begin{table}[b!]
  \begin{center}
    \caption{Energy positions at $k_{<100>}=0$ and effective masses of the four parent bands 
    		of the toy model 
    		used to fit our ARPES data (SI.\ref{SI:Toy-model})
    		and schematically shown 
    		in Figs.~\ref{figS7} and \ref{figS8}.\\
    		}
    \label{tab:table1}
    \begin{tabular}{|c||c|c|} 
    \hline
    & \multicolumn{2}{c|}{\textbf{Parent bands parameters}} \\
      \hline \hline
      & effective mass & energy position \\
      \hline
      \:\: HEB \:\:& $ \:\:\mheb =130 \me$ \:\:& $ \:\:E_\mathrm{HEB}=-8\:\mathrm{meV} \:\:$ \\
      \:\: LEB \:\:& $ \:\:\mleb = 3 \me$ \:\:& $ \:\:E_\mathrm{LEB}=0\:\mathrm{meV} \:\:$ \\
      \:\: HHB \:\:& $ \:\:\mhhb = -50 \me$ \:\:& $ \:\:E_\mathrm{HHB}=9\:\mathrm{meV} \:\:$ \\
      \:\: LHB \:\:& $ \:\:\mlhb = -3 \me$ \:\:& $ \:\:E_\mathrm{LHB}=22\:\mathrm{meV} \:\:$  \\
      \hline
    \end{tabular}
  \end{center}
\end{table}

Interactions between the different bands can be separated in two types.
We assimilate the first type to a Kondo hybridization of strength $\Vhyb \approx 5$~meV,
present in the paramagnetic, HO and AFM states, 
between the heavy electron- (hole-)like band and the light electron- (hole-)band. 
These are the interactions marked with blue and green fonts in panel (b) of Fig.~\ref{figS7}.
We assign the second type to band folding in the ordered states,
induced by the change in the symmetry of the Brillouin zone from BCT in the PM state 
to ST in the HO~\cite{Bareille2014} and AFM states. 
Such band folding implies anti-crossing of electron-like and hole-like bands. 
The corresponding interactions are marked  with black fonts in panel (b) of Fig.~\ref{figS7}.
To accurately fit our ARPES measurements in the HO phase ($x=0$ samples), 
the interaction between the two heavy bands 
in the $\left<100\right>$ direction was found to be $\Vheavy \approx 3$~meV, 
the one between the two light bands was $\Vlight \approx 8$~meV, 
and the interaction between the heavy electron- (hole-)like band 
and the light hole- (electron-)like band was $\Vfold \approx 8.5$~meV. 

From this toy model, the upward shift of the inner electron-like band observed
by ARPES in the AFM phase ($x=0.2$ samples) could result \emph{either} 
from an increased interaction with the heavy electron-like band that lies below, 
\emph{or} from a decreased interaction with the heavy hole-like band that lies above.
These two scenarios would correspond, respectively, to an increase
of $\Vhyb$~from $5$~meV to $8$~meV, or to a decrease of the folding induced
anti-crossing $\Vfold$~from $8.5$~meV to $5.5$~meV.
The interaction parameters between different bands in the two scenarios 
are summarized in Tables~\ref{tab:table2} and~\ref{tab:table3}.
The latter scenario is pictured in the Supplementary Fig.~\ref{figS8}.

\begin{table*}[p!]
  \begin{center}
    \caption{Interaction parameters between the different bands of the toy model 
    		used to fit our ARPES data (SI.\ref{SI:Toy-model}),
    		corresponding to the scenario 1: increased Kondo hybridization when going
    		from the HO to the AFM state.
    		Black-colored values correspond to the bands of URu$_{2-x}$Fe$_{x}$Si$_{2}$ 
    		in both the HO ($x=0.0$) and the AFM ($x=0.2$) phase. 
    		A change in $\Vhyb$ from the blue- to the green-colored value 
    		denotes that the system has passed from the HO to the AFM phase.
    		}
    \label{tab:table2}
    \begin{tabular}{|c||c|c|c|c|} 
    \hline
    & \multicolumn{4}{c|}{\textbf{Interaction parameters -- scenario 1: 
    							  Increased Kondo hybridization
    							  \textcolor{blue}{HO} $\rightarrow$ 
    							  \textcolor{OliveGreen}{AFM}}}\\
      \hline \hline
      & HEB & LEB & HHB & LHB\\
      \hline
      \:\: HEB \:\:& -- & $\Vhyb = $ \textcolor{blue}{5} $\rightarrow$ \textcolor{OliveGreen}{8}~meV & 
      				$\Vheavy = 3$~meV & $\Vfold = 8.5$~meV\\
      \:\: LEB \:\:&  $\Vhyb = $ \textcolor{blue}{5} $\rightarrow$ \textcolor{OliveGreen}{8}~meV & -- & 
      				$\Vfold = 8.5$~meV & $\Vlight = 8$~meV\\
      \:\: HHB \:\:&  $\Vheavy = 3$~meV & $\Vfold = 8.5$~meV & -- & 
      				$\Vhyb = $ \textcolor{blue}{5} $\rightarrow$ \textcolor{OliveGreen}{8}~meV\\
      \:\: LHB \:\:&  $\Vfold = 8.5$~meV & $\Vlight = 8$~meV & 
      				$\Vhyb = $ \textcolor{blue}{5} $\rightarrow$ \textcolor{OliveGreen}{8}~meV & -- \\
      \hline
    \end{tabular}
  \end{center}
\end{table*}

\begin{table*}[t!]
  \begin{center}
    \caption{Interaction parameters between the different bands of the toy model 
    		used to fit our ARPES data (SI.\ref{SI:Toy-model}),
    		corresponding to the scenario 2: decreased folding interaction when going
    		from the HO to the AFM state. This scenario is also shown in Fig.~\ref{figS8}.
    		Black-colored values correspond to the bands of URu$_{2-x}$Fe$_{x}$Si$_{2}$ 
    		in both the HO ($x=0.0$) and the AFM ($x=0.2$) phase. 
    		A change in $\Vfold$ from the blue- to the green-colored value 
    		denotes that the system has passed from the HO to the AFM phase.}
    \label{tab:table3}
    \begin{tabular}{|c||c|c|c|c|} 
    \hline
    & \multicolumn{4}{c|}{\textbf{Interaction parameters -- scenario 2: 
    							  Decreased folding interaction 
    							  \textcolor{blue}{HO} $\rightarrow$ 
    							  \textcolor{OliveGreen}{AFM}}}\\
      \hline \hline
      & HEB & LEB & HHB & LHB\\
      \hline
      \:\: HEB \:\:& -- & $\Vhyb = 5$~meV & $\Vheavy = 3$~meV & 
      					  $\Vfold = $ \textcolor{blue}{8.5} $\rightarrow$ \textcolor{OliveGreen}{5.5}~meV\\
      \:\: LEB \:\:& $\Vhyb = 5$~meV & -- & 
      				 $\Vfold = $ \textcolor{blue}{8.5} $\rightarrow$ \textcolor{OliveGreen}{5.5}~meV & 
      				 $\Vlight = 8$~meV\\
      \:\: HHB \:\:& $\Vheavy = 3$~meV & 
      				 $\Vfold = $ \textcolor{blue}{8.5} $\rightarrow$ \textcolor{OliveGreen}{5.5}~meV 
      				 & -- & $\Vhyb = 5$~meV\\
      \:\: LHB \:\:& $\Vfold = $ \textcolor{blue}{8.5} $\rightarrow$ \textcolor{OliveGreen}{5.5}~meV & 
      				 $\Vlight = 8$~meV & $\Vhyb = 5$~meV & -- \\
      \hline
    \end{tabular}
  \end{center}
\end{table*}

\subsection{\label{SI:ARPES-hv-dependence} Out-of-plane Fermi surfaces of 
UR\lowercase{u}$_{2-x}$F\lowercase{e}$_{x}$S\lowercase{i}$_{2}$ \textit{\lowercase{vs.}} 
UR\lowercase{u}$_{2}$S\lowercase{i}$_{2}$
}

\begin{figure*}[!t]
	\centering
        \includegraphics[clip, width=\textwidth]{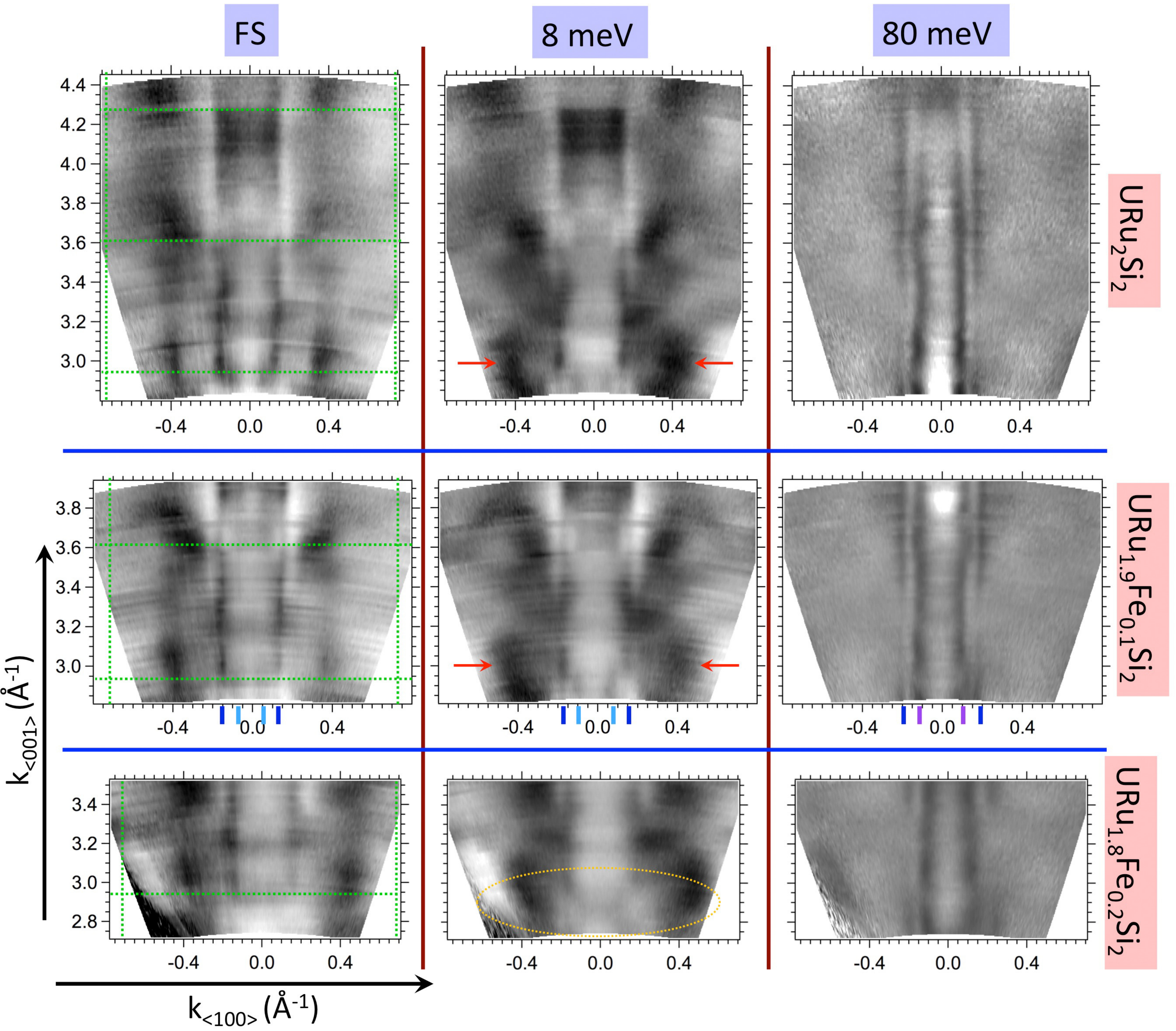}
    	\caption{
        	 Top row: constant energy maps of URu$_{2}$Si$_{2}$ 
        	 along a plan normal to the sample surface and at a binding energy of 0~meV (left), 
        	 8~meV (center) and 80~meV (right).
        	 Middle row: Constant energy maps of URu$_{1.9}$Fe$_{0.1}$Si$_{2}$ 
        	 along a plan normal to the sample surface and at a binding energy of 0~meV (left), 
        	 8~meV (center) and 80~meV (right).
        	 Bottom row: Constant energy maps of URu$_{1.8}$Fe$_{0.2}$Si$_{2}$ 
        	 along a plan normal to the sample surface and at a binding energy of 0~meV (left), 
        	 8~meV (center) and 80~meV (right). 
        	 The dashed green lines denote the borders of the simple tetragonal Brillouin zone. 
        	 Arrows point at some typical closed Fermi contours formed by the heavy bands. 
        	 Dark blue, light blue and purple line segments mark the $k$-location 
        	 of the light-hole band, the M-shaped band and the surface state, respectively. 
        	 The dashed ellipse is around the lower $\Lambda$ point 
        	 where less sharp contours are observed for URu$_{1.8}$Fe$_{0.2}$Si$_{2}$ 
        	 in comparison to other stoichiometries. 
        	 The $k_{x}-k_{z}$ constant energy maps have been acquired 
        	 by varying the photon energy between 20~eV and 67~eV. 
        	 The photon polarisation was linear vertical and the temperature was 1K.
        }
	\label{figS9}
\end{figure*}

Fig.~\ref{figS9} compares the out-of-plane constant energy maps of pure URS and Fe-doped URS at 1K.
Similarly to its in-plane counterpart discussed in the main text, 
the out-of-plane FS of URS presents minimal changes 
as a function of Fe concentration, as shown in Fig.~\ref{figS9} (left column). 
A closer inspection to the constant energy maps at a binding energy of 8~meV (middle column) 
reveals that URu$_{1.8}$Fe$_{0.2}$Si$_{2}$ shows less sharp contours at the lower $\Lambda$ point 
($k_{<001>}$ = $2.95$~\AA$^{-1}$, dashed ellipse) with respect to other Fe concentrations. 
This observation is in line with the data presented in the main text, 
where the near-$E_{F}$ electronic structure 
disperses down to a smaller binding energy at $\Lambda$ for $x$ = $0.2$. 
Although the light-hole conduction band and the M-shaped feature 
show no appreciable out-of-plane dispersion,
we cannot assign a surface origin to them because the corresponding
in-plane contours shown in the main text, Figs.~2(a-d), 
are in excellent agreement with bulk LSDA calculations~\cite{Elgazzar2009,Oppeneer2010}. 
We underline that the limited $k_z$  resolution of the experimental technique
(i.e. inversely proportional to the photoelectron escape depth~\cite{Strocov2003}, 
which is almost as high as one-fourth of the simple tetragonal BZ in our case) 
can sometimes mask the out-of-plane dispersion in an ARPES experiment.

The middle column of Fig.~\ref{figS9} reveals the existence of closed constant energy contours 
(some of them marked by arrows) in the $k_{<100>} - k_{<001>}$ plane: 
a proof for non-negligible out-of-plane dispersion of the heavy bands, hence of their 3D bulk character. 
The out-of-plane dispersion of the heavy bands is also captured 
by the supplementary ancillary movie.
showing the $E-k_{<100>}$ dispersion for successive $k_{<001>}$ values. 
On the other hand, the light-hole conduction band (dark blue marker), 
the M-shaped band (light blue marker) and the surface state (purple marker) 
show no appreciable dispersion along $k_{<001>}$.

\subsection{\label{SI:PM-AFM} Changes in the electronic structure of  
UR\lowercase{u}$_{2-x}$F\lowercase{e}$_{x}$S\lowercase{i}$_{2}$
across the PM/AFM transition}

\begin{figure*}[!p]
	\centering
        \includegraphics[clip, width=0.9\textwidth]{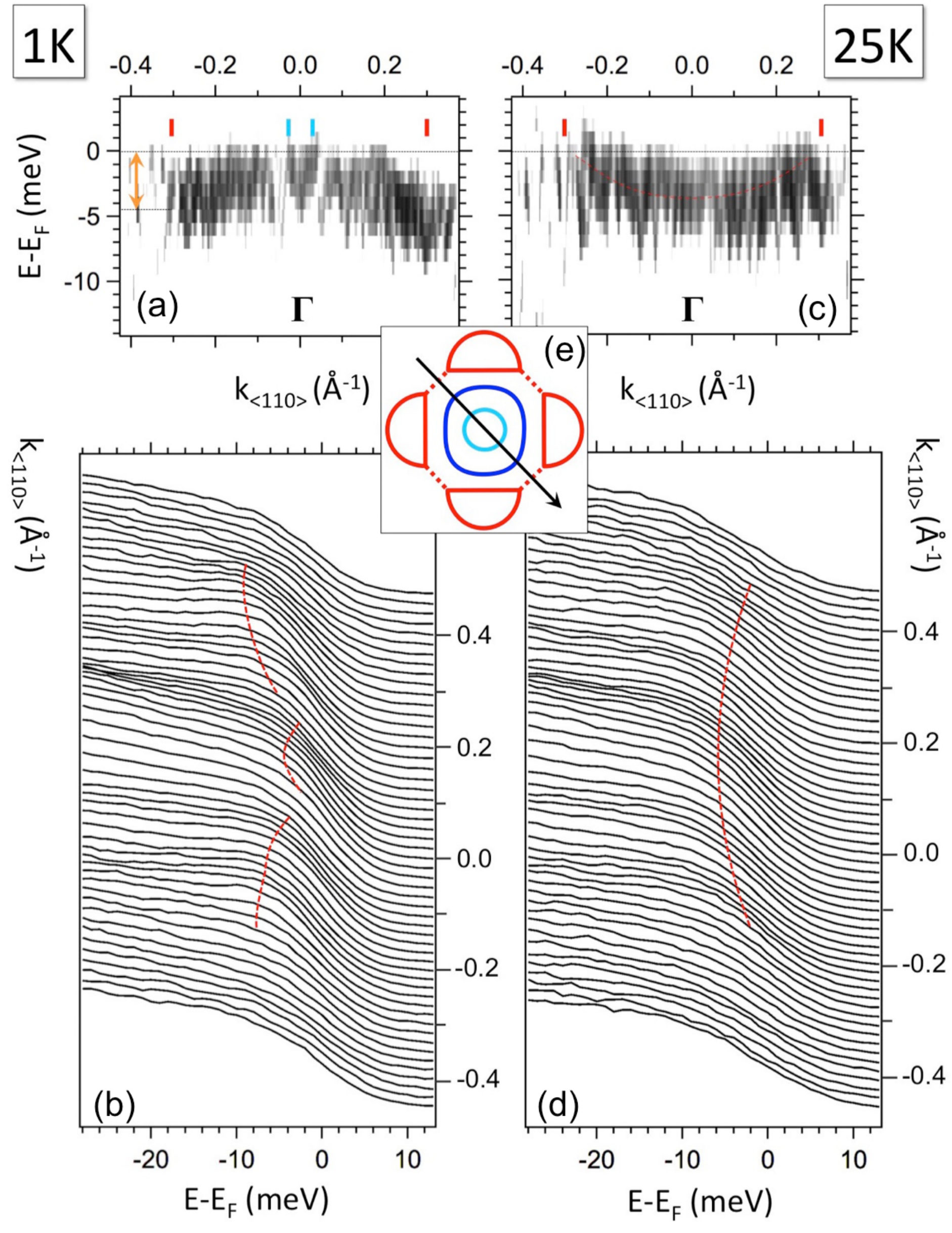}
   		\caption{
        	 (a,~b)~ARPES energy-momentum maps (2D curvature~\cite{Zhang2011}) 
        	 along $k_{<110>}$ at 1K in the AFM phase and 25K in the PM phase, 
        	 respectively. 
        	 The blue bars show the Fermi momenta of the 
        	 shallow heavy electron-like pocket around $\Gamma$ in the AFM state,
        	 while the red bars indicate the Fermi momenta of the large diamond-like
        	 Fermi surface in the PM state, which opens a gap of about 5~meV 
        	 in the AFM state (orange arrow).
        	 (c,~d) Corresponding raw ARPES energy distribution curves.
        	 Red markers are guides to the eyes
        	 showing, in the AFM state, the M-shaped band forming the heavy electron pocket 
        	 and hole-like band crossing $E_F$ around $\Gamma$,
        	 and the large electron pocket forming a diamond in the PM state.
        	 (e)~Schematic representation of the Fermi surfaces observed
        	 in the AFM phase. The arrow shows the direction of measurement
        	 in this figure. The dotted lines schematize the large diamond-like
        	 Fermi surface in the PM state~\cite{Bareille2014}.
        }
	\label{figS10}
\end{figure*}

\begin{figure*}[!p]
	\centering
        \includegraphics[clip, width=0.8\textwidth]{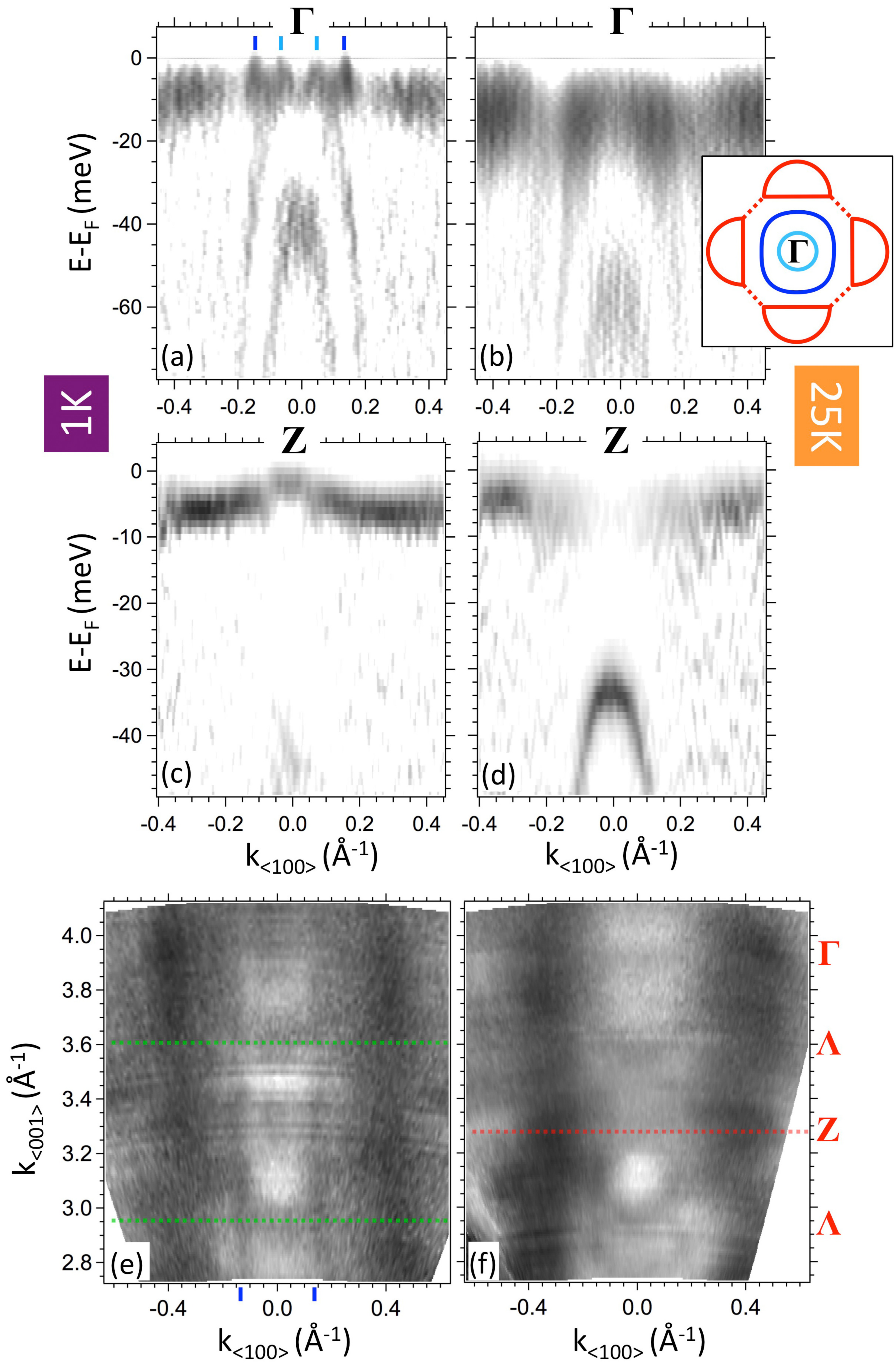}
    	\caption{
        	 (a-d) Near-$E_F$ energy dispersion of URu$_{1.8}$Fe$_{0.2}$Si$_{2}$ 
        	 at 1~K (AFM phase, left) and at 25 K (PM phase, right). 
        	 The electronic band structure around $\Gamma$ and Z shows no signs 
        	 of Ru~$4d$~-~U~$5f$ hybridisation at 25~K.
        	 (e,~f) Out-of-plane Fermi surface of URu$_{1.8}$Fe$_{0.2}$Si$_{2}$ at 1~K 
        	 (AFM phase, left) and at 25~K (PM phase, right). 
        	 At 25 K there are no fingerprints of the cylindrical Fermi sheet. 
        	 Line segments in panels (a) and (e) correspond to the contours of the inset. 
        	 Dotted lines in panels (e) and (f) denote the borders of the 
        	 ST and the BCT Brillouin zones, respectively. 
        	 Data around $\Gamma$ and Z have been acquired using 50.5~eV and 31~eV photons, 
        	 respectively. The out-of-plane Fermi surfaces have been acquired 
        	 by varying the photon energy from 20~eV to 57~eV.
        	 The photon polarisation was linear horizontal. 
        	 In order to enhance the experimental features in panels (a-d), 
        	 the 2D curvature of the ARPES intensity is presented \cite{Zhang2011}.
        }
	\label{figS11}
\end{figure*}

Similar to the pure URu$_2$Si$_2$, whose in-plane Fermi-surface in the PM state
gets gapped along the $<110>$ directions in the HO state~\cite{Bareille2014},
the electronic structure of URu$_{1.8}$Fe$_{0.2}$Si$_{2}$ also develops a gap
along the $<110>$ directions upon the PM$\rightarrow$AFM phase transition.
Figs.~\ref{figS10} and \ref{figS11} present the changes in the electronic structure 
of URu$_{2-x}$Fe$_{x}$Si$_{2}$, respectively along $k_{<110>}$ and $k_{<100>}$,
across the PM/AFM phase transition.

Fig.~\ref{figS10} shows the electronic dispersion along $k_{<110>}$, 
of URu$_{1.8}$Fe$_{0.2}$Si$_{2}$ at 1K (left column, AFM phase) 
and 25K (right column, PM phase).
In the PM state, the most prominent feature is a large
heavy electron band crossing $E_F$, forming a diamond-like Fermi surface
centered around $\Gamma$.
We note that this heavy electron band is qualitatively captured by the parent bands 
of the toy model presented in Fig.~\ref{figS7}.
In the AFM phase, the band structure undergoes substantial changes:
a gap of about 5~meV opens along $k_{<110>}$ in the diamond Fermi surface,
at the so-called ``hot-spots'', and the band dispersion
becomes M-shaped, forming a shallow heavy electron pocket 
surrounded by a heavy hole-like pocket around $\Gamma$.

Fig.~\ref{figS11} compares the electronic dispersion along $k_{<100>}$, 
and corresponding out-of-plane FS contours, 
of URu$_{1.8}$Fe$_{0.2}$Si$_{2}$ at 1K (left column, AFM phase) and 25K (right column, PM phase). 
In the PM state, the heavy bands near $E_F$ show again the large electron-like pocket
forming the diamond-like Fermi surface.
Deep into the AFM state, as also shown in the main text, such electron pocket
transforms into an M-shaped band, giving rise to four off-centered
``Fermi petals'' at $k_{<100>} \approx 0.4$~\AA$^{-1}$.
All the changes in the near-$E_F$ electronic structure across the PM/AFM transition
are similar to the ones observed in pure URu$_2$Si$_2$ 
across the PM/HO transition~\cite{Bareille2014}, as also discussed in the main text.

We note that changes between the AFM and the PM phase 
as shown in Figs.~\ref{figS10} and \ref{figS11}
are fundamental and not related to extrinsic factors such as surface quality.
For instance, the surface quality in Fig.~\ref{figS11}(d) was higher than in Fig.~\ref{figS11}(c) 
as can be inferred from the well-resolved surface state in the former.
The differences in the near-$E_{F}$ dispersion of the light-hole conduction band 
are also reflected in the out-of-plane FS contours, bottom panels of Fig.~\ref{figS11}. 
In the AFM phase [Fig.~\ref{figS11}(e)], one can clearly track Fermi-surface contours 
reflecting the apparent lack of $k_{<001>}$ dispersion of the light-hole band.
These open FS contours are barely visible in the PM phase [Fig.~\ref{figS11}(f)]. 
This is not surprising knowing that the light-hole conduction band exhibits no intensity 
at energies higher than the $f$-band continuum [Fig.~\ref{figS11}(b)]. 
As a side note, we point out that the apparent closed contours 
in the vicinity of the lower $\Lambda$ point in Fig.~\ref{figS11}(f) 
do not belong to a single state: they consist of small portions of the light-hole conduction band 
(outer part of the contour) and a shallow electron pocket (inner part of the contour). 
The shallow electron pocket at $\Lambda$ -also known as the `lentil' \cite{Bareille2014}- 
has been already discussed in the main text.



\clearpage

%

\end{document}